\documentclass[sigconf]{acmart}

\setcopyright{none}
\renewcommand\footnotetextcopyrightpermission[1]{}

\usepackage{tikz}
\usepackage{algorithm}
\usepackage{algpseudocode}
\usepackage{subcaption}
\usepackage{booktabs}
\usepackage{enumitem}
\usepackage{array}
\usepackage{listings}
\usepackage{graphicx}
\usepackage{tabularx}
\usepackage{longtable}
\usepackage{placeins}
\usepackage{float}

\lstdefinestyle{json}{
    basicstyle=\ttfamily\small,
    breaklines=true,
    breakatwhitespace=true,
    frame=single,
    framesep=4pt,
    xleftmargin=2pt,
    xrightmargin=2pt,
    rulecolor=\color{black!25},
    backgroundcolor=\color{black!3},
    numbers=none,
    showstringspaces=false,
    keepspaces=true,
    tabsize=2,
    moredelim=[s][\color{blue!45!black}]{"}{"},
    literate=
      {:}{{{\bfseries :}}}{1}
      {\{}{{{\bfseries \{}}}{1}
      {\}}{{{\bfseries \}}}}{1}
      {[}{{{\bfseries [}}}{1}
      {]}{{{\bfseries ]}}}{1},
}

\begin{document}

\title{Multi-Objective Scheduling of LLM-Based Multi-Agent Workflows in Cloud Computing}

\author{Goshgar Can Ismayilov}
\affiliation{%
  \city{Istanbul}
  \country{Turkey}}
\email{goshgar.ismayilov@boun.edu.tr}

\begin{abstract}
Large language models (LLMs) have been increasingly composed into multi-agent workflows to accomplish complex and multi-step tasks. As the open-weight models continue to close the performance gap with the proprietary ones, model parameters are becoming a less significant source of competitive advantage in the artificial intelligence market. Yet how these models are efficiently scheduled onto distributed infrastructure such as the cloud, with respect to overall latency, monetary cost and resource utilization, remains largely unaddressed in the literature despite being an increasingly critical problem. Existing agentic workflow frameworks mostly focus on output quality and task success rates over popular benchmarks, rarely accounting for execution dynamics of LLM-based agents such as inference latency, computation power of heterogeneous resources, cloud-specific billing period and resource provisioning overheads. Therefore, this paper formulates multi-agent scheduling in cloud computing as a multi-objective optimization problem that jointly minimizes makespan and cost while maximizing resource utilization. Building on this formulation, we propose MO-SAIC (\underline{M}ulti-\underline{O}bjective \underline{S}cheduling of Multi-\underline{A}gent Workflows \underline{i}n \underline{C}loud Computing), a scheduling framework comprising four multi-objective schedulers with Non-Dominated Sorting Genetic Algorithm II (NSGA-II), Weighted-Sum Genetic Algorithm (W-GA), a multi-objective adaptation of Heterogeneous Earliest Finish Time (MO-HEFT) and a random baseline (RAND). We evaluate the schedulers from the workflow scheduling, multi-objective optimization and system perspectives to measure their overall performance, quality of their Pareto-optimal fronts and impacts of workflow topology, cloud resource composition and scalability parameters. We report the experimental results with their statistical significance via the Wilcoxon test. Lastly, we discuss the limitations of MO-SAIC as well as future research directions. 
\end{abstract}


\maketitle
\pagestyle{plain}

\section{Introduction}
\label{sec:introduction}

Consider an analogy from the history of cryptography. The field spent its early years as a state secret where advantage depended on which government could keep the cryptographic techniques hidden the longest through the mid-twentieth century. The decades since have reversed this trend, especially with the emergence of public-key cryptography. After that shift, security came to rest not on hiding the techniques but on protecting comparatively small secret keys. This transition follows the famous Kerckhoffs's principle~\cite{kerckhoffs1883}, stating that a cryptosystem should not rely on the secrecy of the algorithm. We believe that large language models and agentic systems are currently heading toward an analogous transition. As open-weight models continue to close the performance gap with proprietary ones and model architectures are increasingly shared, model parameters are becoming a less significant source of competitive advantage. What increasingly differentiates is becoming not which models are used but how they are used at scale with respect to the efficient management of overall latency, cost and resource utilization. Rather than the model layer of artificial intelligence, the operational layer will be the harder challenge for large organizations, making the scheduling of agentic workflows on distributed infrastructures an increasingly critical research problem.

LLMs have rapidly progressed from static text generators to autonomous agents with advanced capabilities of planning, reasoning, invoking external tools and acting on their own. Further techniques for LLMs including tool-augmented generation \cite{schick2023} and interleaving of reasoning and acting \cite{yao2022} have enabled these agents to actively query APIs, execute code and retrieve information to complete complex tasks. They often chain together many such steps instead of acting on a single request in isolation. This conceptual shift in LLMs has also been documented across a rapidly growing body of survey literature \cite{wang2024}. However, all these autonomous actions consume real computational resources in the background, incurring latency and cost. As agents are integrated into production systems at increasing scale, their total operational footprint has become a significant concern to address in its own right. 

A single LLM-based agent is rarely sufficient for complex and multi-step tasks, which has led to composing multiple specialized agents into multi-agent workflows. An increasing number of works in the literature address the automation and optimization of this multi-agent workflow composition. For instance, AFlow \cite{aflow} employs a Monte Carlo tree search technique to iteratively sample, evaluate and refine different workflows. EvoFlow \cite{evoflow} instead evolves a diverse population of workflows based on an evolutionary algorithm with niching solution selection. However, once workflow topology has been fixed, no prior work has sufficiently addressed how its agents should be effectively assigned to and executed on physical cloud resources. In addition, while scheduling a single-agent system is relatively straightforward, the same is not true for multi-agent systems since such systems allow distinct agents to run on entirely separate resources. This gap motivates the problem of multi-agent workflow scheduling in distributed systems \cite{mawcs}.

\begin{figure*}[htbp]
\centerline{\includegraphics[width=0.99\textwidth]{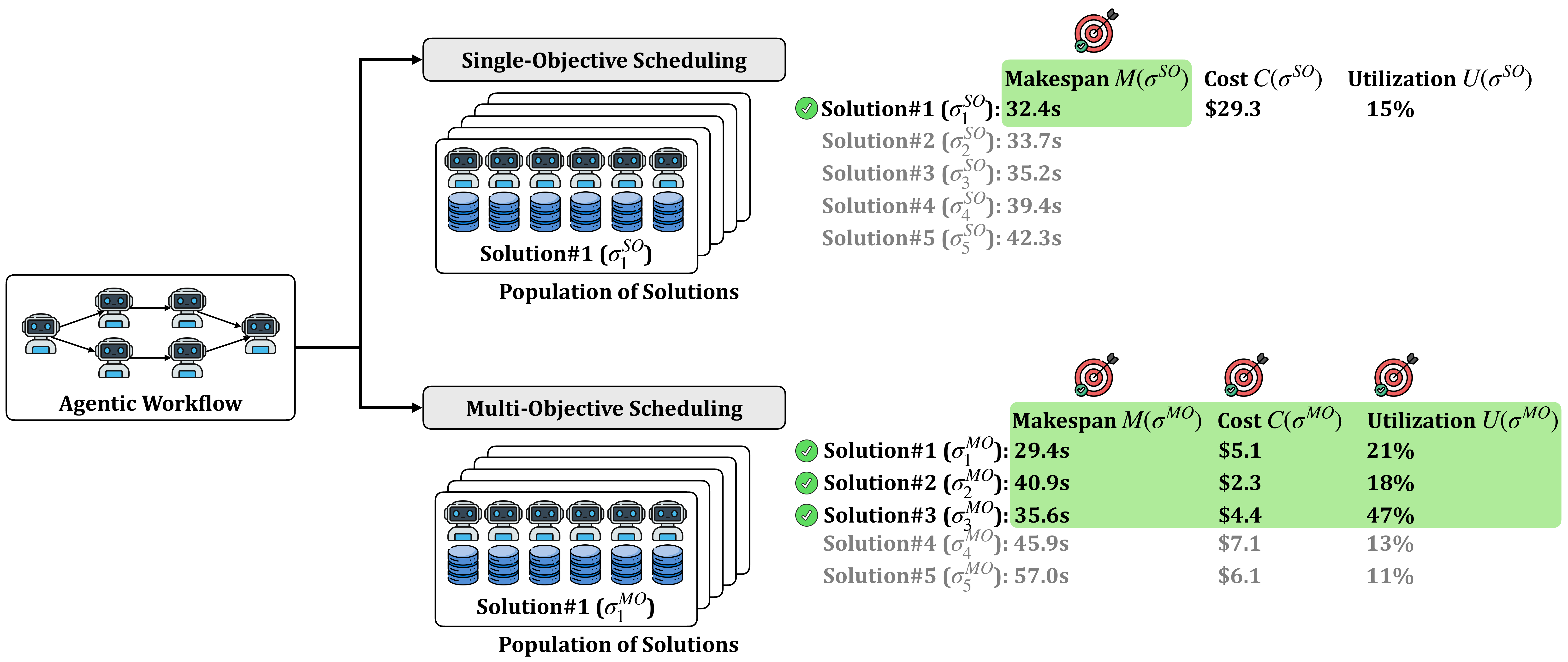}}
\caption{Single-objective scheduling commits to a single solution purely via makespan (top) while multi-objective scheduling selects multiple solutions with distinct trade-offs across makespan, cost and utilization (bottom).}
\label{fig:intro}
\end{figure*}

Scheduling traditional workflows across a pool of resources is not a new problem in the literature. Multi-objective evolutionary algorithms, most notably NSGA-II \cite{deb2002}, have provided a technique to search for Pareto-optimal trade-offs when two or more objectives (e.g., makespan and cost) cannot be simultaneously minimized. In this respect, NSGA-II has been applied extensively to workflow scheduling in cloud computing \cite{durillo2012, ismayilov2020} for the optimization of several objectives including makespan, cost, utilization, energy consumption, reliability and degree of imbalance. However, existing scheduling techniques are primarily designed for traditional workflows (e.g., Montage in astronomy \cite{deelman2015}), not for agentic workflows where they do not consider the inherent characteristics of agents (e.g., inference time, tool calling time, output token size). Additionally, agents increasingly require heterogeneous CPU and GPU instances rather than a CPU-only pool. Figure~\ref{fig:intro} illustrates the distinction between the single-objective and multi-objective scheduling where the single-objective scheduler commits to a single solution purely via makespan without evaluating potential trade-offs even when it is extremely costly. The multi-objective scheduler instead offers multiple solutions that minimize latency, minimize cost or maximize resource utilization. This clearly shows that multi-agent workflow scheduling is inherently a multi-objective problem. Therefore, we position our work precisely at this intersection.

\subsection{Contributions}

This paper makes the following contributions to the literature:

\begin{itemize}
    \item \textbf{A formal model of multi-agent workflow scheduling in cloud computing.} We formulate multi-agent workflow scheduling in cloud computing as a multi-objective optimization problem by considering the minimization of workflow makespan and monetary cost and the maximization of resource utilization. This problem abstracts four critical models: (i) a LLM-based agent model with inference latency, model-specific throughput, tool latency and output token size; (ii) a multi-agent workflow model through  canonical agentic reasoning patterns including chain-of-thought, self-consistency, tree-of-thought, debate and self-refine; (iii) a heterogeneous cloud resource model by unifying CPU and GPU computation capacity; and (iv) a workflow simulation model to execute workflows on resources by considering cloud-specific issues such as hourly-billing and resource (de)provisioning overheads.

    \item \textbf{MO-SAIC -- a multi-objective scheduling framework.} We propose a novel multi-objective multi-agent workflow scheduling framework for cloud computing, MO-SAIC, with the total of four schedulers: (i) a non-dominated sorting genetic algorithm scheduler (NSGA-II) as Pareto metaheuristic; (ii) a weighted-sum genetic algorithm scheduler (W-GA) as a scalarized metaheuristic; (iii) a multi-objective HEFT scheduler (MO-HEFT) as a classical list heuristic; and (iv) a random baseline scheduler (RAND) with no intelligence.
    
    \item \textbf{A controllable synthetic workflow generator.} We employ a parametric and synthetic multi-agent workflow generator \cite{mawge} to generate workflows with a diverse combination of canonical agentic reasoning patterns with given workflow size, depth and width. This generator allows controlled empirical study of how workflow topology affects scheduling decisions and quality.

    \item \textbf{Extensive experimental study.} We extensively evaluate these four schedulers under various test scenarios to measure: (i) their overall performance relative to a random baseline scheduler; (ii) quality of their Pareto-optimal fronts; (iii) effects of workflow topology with changing size, depth and width; and (iv) effects of cloud resources with resource size and CPU/GPU pool composition. In our experiments, we use several performance measures from three distinct perspectives: (i) optimization objectives, resource load imbalance, distinct resource count for the workflow scheduling perspective; (ii) number of non-dominated solutions, Schott's spacing, hypervolume and visual Pareto-optimal fronts for the multi-objective perspective and (iii) scheduling runtime for the system perspective. The statistical significance is shown via Wilcoxon rank-sum test across independent runs on distinct workflows.

    \item \textbf{A further discussion of limitations and future directions.} We also discuss the current assumptions and limitations of MO-SAIC against the practical requirements for production deployments and the potential research directions to extend this research.

\end{itemize}

\subsection{Organization}

The remainder of this paper is organized as follows. Section~\ref{sec:related-work} reviews the literature of agentic workflows, traditional workflow scheduling and multi-objective optimization, by highlighting the current gap in the literature. Section~\ref{sec:problem} formally defines the problem as a multi-objective optimization problem, by specifying the optimization objectives and constraints as well as the agent, workflow, cloud and simulation models. Section~\ref{sec:framework} presents our multi-objective scheduling framework with four schedulers of NSGA-II, W-GA, MO-HEFT and RAND. Section~\ref{sec:experimental-setup} describes our experimental methodology for the benchmark workflows, performance measures and statistical methodology to compare the schedulers. Section~\ref{sec:experimental-results} reports the experimental results and observations under various test scenarios. Section~\ref{sec:discussions} discusses the current limitations and future directions while Section~\ref{sec:conclusion} concludes the paper.

\section{Related Work}
\label{sec:related-work}

This section reviews the relevant works from the literature in three main perspectives including LLM-based agentic workflows, multi-objective workflow scheduling in cloud computing and agent scheduling for LLM serving. 

\subsection{LLM-Based Agentic Workflows}

Many works in the literature have built LLM-based agentic workflows by composing multiple agents with heterogeneous roles and models. Wu et al. \cite{wu2023} present AutoGen, a general framework where multiple agents with different roles and tools communicate with one another to accomplish a common task collaboratively. Hong et al. \cite{hong2024} propose MetaGPT which employs standardized operating procedures from human software-engineering workflows for role-specialized agents. This enables agents to have human-like expertise and verify intermediate results to reduce error. Li et al. \cite{li2023} design Camel to study how two agents autonomously and complementarily can cooperate toward a common goal with role-playing. Park et al.~\cite{park2023} populate an interactive sandbox environment with 25 generative agents that plan, act and reflect over time to produce emergent social behavior (e.g., planning their days, sharing news). For architectural characterization, construction techniques and application domains, Wang et al. \cite{wang2024} present broader perspectives.

The multi-agent workflows all these frameworks orchestrate are mostly composed from a small set of canonical reasoning patterns. For instance, Wei et al. \cite{wei2022} show that eliciting intermediate reasoning states via chain-of-thought substantially improves LLM performance on complex tasks. Wang et al. \cite{wang2022} extend the concept by sampling multiple independent reasoning paths and aggregating into a single agent via self-consistency. Yao et al. \cite{yao2023} address explicit search over a tree of intermediate reasoning states rather than a linear reasoning. Du et al. \cite{du2023} show that two agents debating and critiquing the answers of the other over several rounds improves factuality and reasoning. Madaan et al.~\cite{madaan2023} show that only a single agent iteratively critiquing its own response improves the response quality over several rounds. Finally, Yao et al. \cite{yao2022} propose ReAct, which interleaves reasoning traces with tool actions so that agents dynamically revise their reasoning based on their observations.

\subsection{Multi-Objective Workflow Scheduling in Cloud Computing}

Multi-objective optimization has been extensively applied to workflow scheduling in cloud computing, most commonly targeting the trade-off between makespan and monetary cost. Durillo et al. \cite{durillo2012} adapt the classical HEFT heuristic \cite{topcuoglu1999} into MO-HEFT which is a Pareto-based list scheduling algorithm to maintain a set of partial scheduling solutions and measure its performance against synthetic and real-world workflows. Zhu et al. \cite{zhu2015} instead propose a NSGA-II-based evolutionary multi-objective metaheuristic for the joint optimization of the makespan and cost. Li et al. \cite{li2021} later improve the search performance of NSGA-II \cite{deb2002} for workflow scheduling through a scoring criterion and a dynamic hierarchical population. Ismayilov et al. \cite{ismayilov2018, ismayilov2020} extend multi-objective workflow scheduling to a dynamic setting so that it reacts better to environmental changes such as resource failures by incorporating a neural network into NSGA-II. They consider makespan, cost, utilization, energy, reliability and degree of imbalance as objectives.  

To minimize energy consumption together with makespan and cost, Yassa et al.~\cite{yassa2013} combine a hybrid particle swarm optimization with the dynamic voltage and frequency scaling technique. This enables the underlying processors to operate in different voltage supply levels by changing their clock frequencies. Later, Verma et al. \cite{verma2017} hybridize particle swarm optimization with a budget and deadline-constrained HEFT especially for scientific workflows. Mohammadzadeh et al. \cite{mohammadzadeh2023} propose another hybrid algorithm based on the seagull and grasshopper optimization algorithms to jointly optimize makespan, cost, energy and throughput, now with resources from multiple cloud providers. Abazari et al. \cite{abazari2019} broaden the set of objectives further to include the minimization of security risk in the cloud and the response time. 

\subsection{Model and Agent Scheduling for LLM Serving}

There exists a smaller and more recent group of works for scheduling of LLM executions across LLM engines. Luo et al. \cite{luo2025} propose Autellix which enriches schedulers with program-level statistics and minimizes end-to-end latency. Lin et al. \cite{lin2024} consider the data dependencies between consecutive LLM calls through semantic variables so that they can be co-scheduled. Jiang et al. \cite{jiang2023} formally define the scheduling problem of LLM serving over a set of heterogeneous GPU resources as a constrained optimization problem. Griggs et al. \cite{griggs2024} assign LLM calls to GPU resources with distinct prices and capabilities to reduce cost under a certain latency target. Ong et al. \cite{ong2025} focus on routing to two classes of models as strong models generating high-quality responses at a higher cost and weak models generating lower-quality responses but at a reduced cost. They present a training framework that dynamically select between these models during inference. Finally, the recent work \cite{peng2026} proposes HexAGenT as a workflow and heterogeneity-aware scheduling algorithm for the online workflow scheduling problem. Unlike these works, our focus is offline scheduling that assigns each agent to a cloud resource and executes the model on that instance while optimizing makespan, cost and utilization jointly.

\section{Multi-Objective Multi-Agent Workflow Scheduling}
\label{sec:problem}

Formulating the multi-agent workflow scheduling problem requires the characterization of (i) the individual LLM-based agents, (ii) the agentic workflows as the composition of individual agents with a certain topology, (iii) the cloud computing as the infrastructure provider to run the workflows and (iv) the simulator to simulate the execution of the workflows in the cloud under certain conditions and constraints. Therefore, this section develops these models to culminate in a formal multi-objective optimization of the problem to be addressed. Refer to Table~\ref{table:notation} for the mathematical notations used throughout this section and the remainder of the paper.

\begin{table}[htbp]
\centering
\caption{Descriptions of common mathematical notations used throughout this paper.}
\label{table:notation}
\begin{tabular}{ll}
\toprule
\textbf{Notation} & \textbf{Description} \\
\midrule
$\alpha \in \mathcal{A}$ & An agent in the finite agent set $\mathcal{A}$ \\
$r \in \mathcal{R}$ & A cloud resource in the resource pool $\mathcal{R}$ \\
$\mathrm{cp}(r)$ & Unified compute power of resource $r$ \\
\midrule
$X$ & Decision space of resource assignments \\
$\sigma$ & A scheduling solution \\
$\mathcal{P}$ & A population of candidate solutions \\
$M(\sigma)$, $C(\sigma)$, $U(\sigma)$ & Makespan, cost, utilization of $\sigma$ \\
$\sigma' \prec \sigma$ & $\sigma'$ Pareto-dominates $\sigma$ \\
$\sigma' \preceq \sigma$ & $\sigma'$ weakly dominates $\sigma$ \\
$F_0(\mathcal{P})$ & Pareto-optimal front of $\mathcal{P}$ \\
$R(\sigma)$ & Set of cloud resources $\sigma$ assigns agentic tasks \\
\midrule
$\hat{f}(\sigma)$ & Min-max normalized objectives of $\sigma$ in $[0,1]^3$ \\
$d_i$ & Nearest-neighbor distance of front member $i$ \\
$\mathrm{SS}(F_0)$ & Schott's spacing of front $F_0$ \\
$\Lambda(\cdot)$ & Lebesgue measure (volume) \\
$\mathbf{r} = (r_1,r_2,r_3)$ & Hypervolume reference point \\
$\varepsilon$ & Hypervolume reference margin \\
$\mathrm{HV}(F_0)$ & Hypervolume of front $F_0$ \\
\bottomrule
\end{tabular}
\end{table}

\subsection{LLM-Based Agent Model}

An agent is an LLM-driven computational system that acts autonomously through its own sequence of actions including inference and tool calling. Each agent in an agentic workflow $\alpha \in \mathcal{A}$ is assigned to a model $m(\alpha) \in \mathcal{M}$ from the space of available LLMs $\mathcal{M}$. Every model $m \in \mathcal{M}$ is characterized by a time-to-first-token (TTFT) latency $\phi_m$ (seconds) and a token generation throughput $\rho_m$ (tokens/second). Given an output size $o(\alpha) \in \mathbb{Z}_{>0}$ (i.e., the number of tokens $\alpha$ generates as response), its inference time $L_{\mathrm{inf}}(\alpha)$ can be computed as:
\[
L_{\mathrm{inf}}(\alpha) \;=\; \phi_{m(\alpha)} \;+\; \frac{o(\alpha)}{\rho_{m(\alpha)}}
\]
where it adds the time to stream the output after the initial token generation latency. During its execution, an agent might invoke external tool calls (e.g., for web search) from the finite set of tools $\Theta(\alpha) = \{\theta_1, \ldots, \theta_k\}$. Each tool $\theta_i$ is defined as:
\[
\theta \;=\; (\ell_\theta,\, n_\theta)
\]
where $\ell_\theta \in \mathbb{R}_{>0}$ is the per-invocation latency and $n_\theta \in \mathbb{Z}_{>0}$ is the number of times $\alpha$ invokes it. Total tool latency is computed as:
\[
L_{\mathrm{tool}}(\alpha) \;=\; \sum_{\theta \in \Theta(\alpha)} \ell_\theta \cdot n_\theta
\]
where we assume the invocations as sequential and non-overlapping. In our experimental setting, tools are classified into two groups as the internal tools with low-latency and the external tools with relatively high-latency. Refer to Table~\ref{tab:model_performance} for the specifications of the models considered in this work. Note that this catalog serves solely to parameterize $\phi_m$ and $\rho_m$ with values for our simulation and our agent model itself remains independent of this particular set of models.

\subsection{Multi-Agent Workflow Model}
\label{sec:workflow-model}

A multi-agent workflow is a directed acyclic graph (DAG) $G = (\mathcal{A}, E)$, where $\mathcal{A}$ is the finite set of agents and $E \subseteq \mathcal{A} \times \mathcal{A}$ is a set of directed edges. Each edge $(\alpha, \alpha') \in E$ indicates that agent $\alpha'$ depends on the output of agent $\alpha$ and hence, $\alpha$ must complete before $\alpha'$ can start execution. For an agent $\alpha \in \mathcal{A}$, we write $\mathrm{pred}(\alpha) = \{\alpha' \in \mathcal{A} : (\alpha', \alpha) \in E\}$ and $\mathrm{succ}(\alpha) = \{\alpha' \in \mathcal{A} : (\alpha, \alpha') \in E\}$ to define predecessors and successors, respectively. The workflows we generate are not arbitrary DAGs. Instead, they are composed of several canonical agentic reasoning patterns.

\begin{figure*}[htbp]
    \centering
    \begin{subfigure}[b]{0.33\textwidth}
        \includegraphics[width=\textwidth]{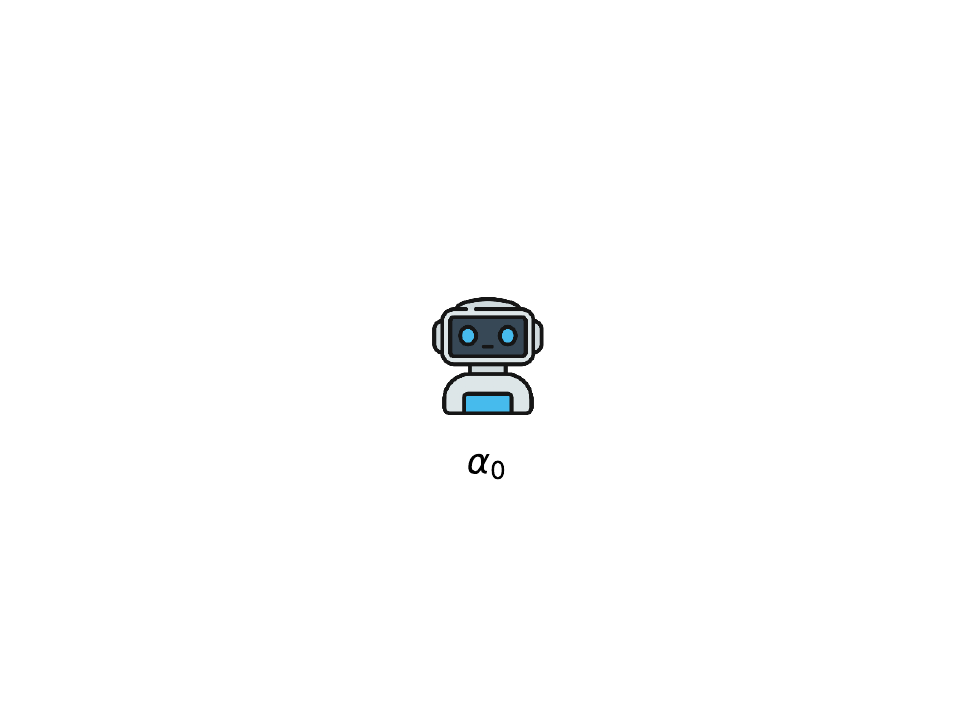}
        \caption{Single-Agent (SA).}
        \label{fig:pattern-sa}
    \end{subfigure}
    \begin{subfigure}[b]{0.33\textwidth}
        \includegraphics[width=\textwidth]{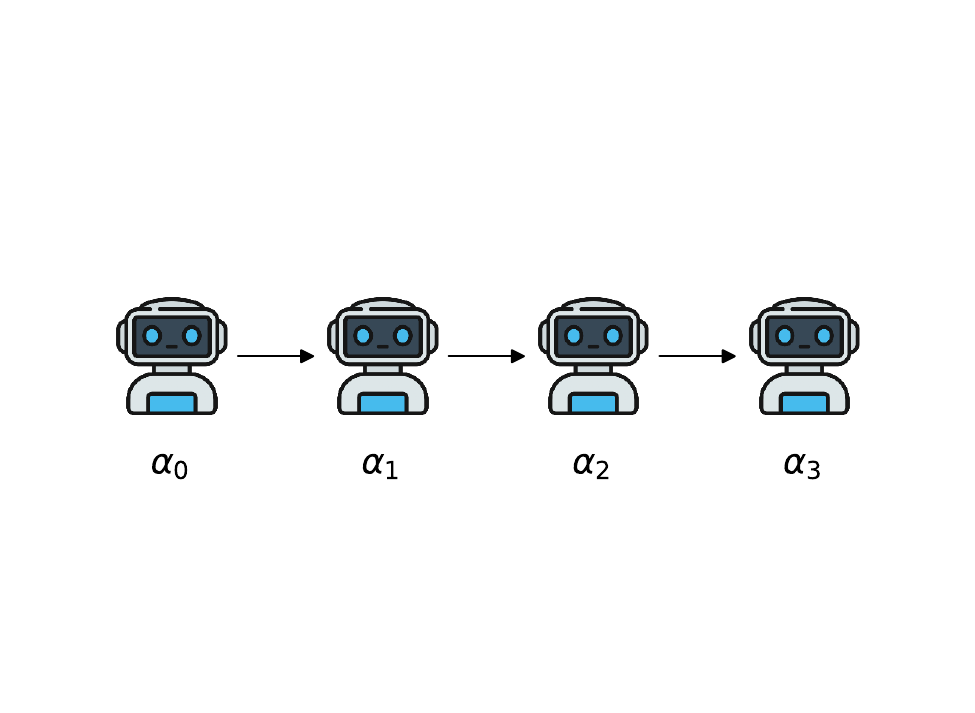}
        \caption{Chain-of-Thought (CoT).}
        \label{fig:pattern-cot}
    \end{subfigure}
    \begin{subfigure}[b]{0.33\textwidth}
        \includegraphics[width=\textwidth]{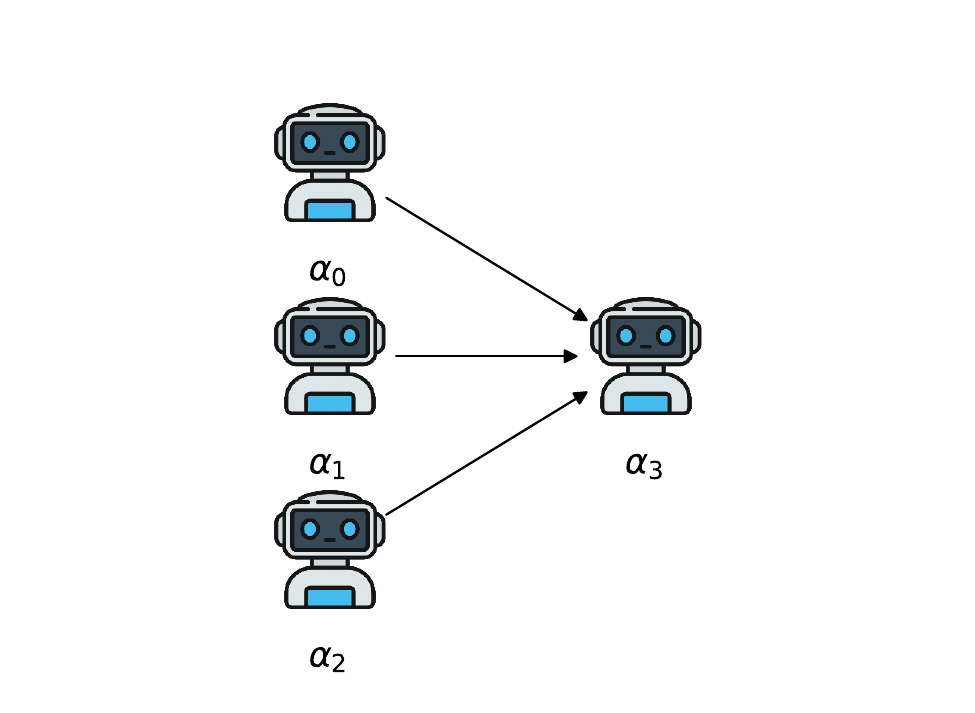}
        \caption{Self-Consistency (SC).}
        \label{fig:pattern-sc}
    \end{subfigure} \\
    \begin{subfigure}[b]{0.33\textwidth}
        \includegraphics[width=\textwidth]{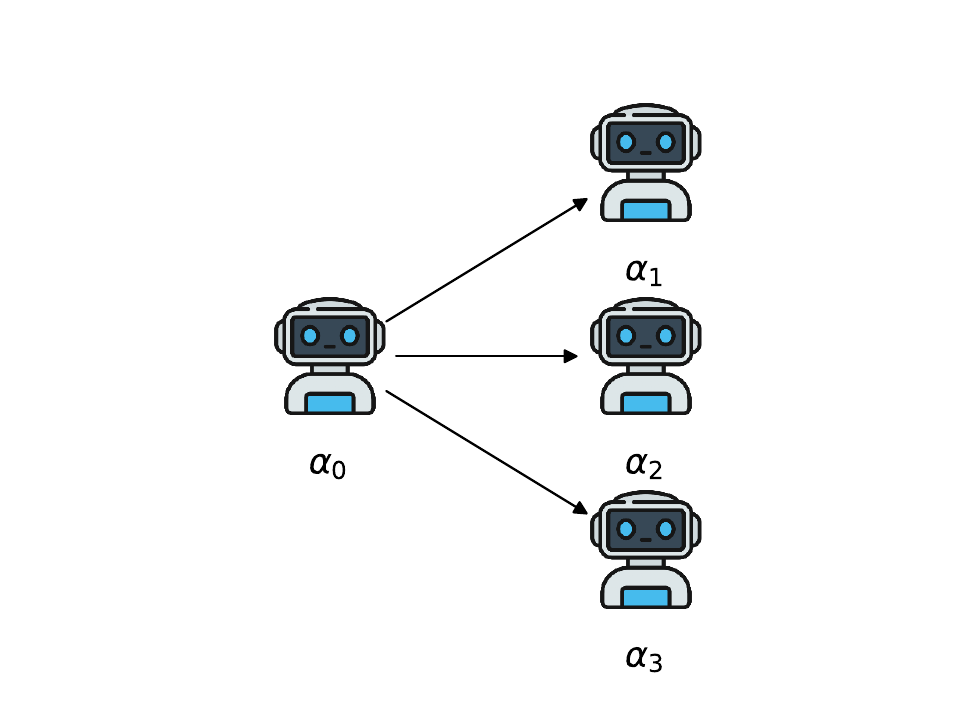}
        \caption{Tree-of-Thought (ToT).}
        \label{fig:pattern-tot}
    \end{subfigure}
    \begin{subfigure}[b]{0.33\textwidth}
        \includegraphics[width=\textwidth]{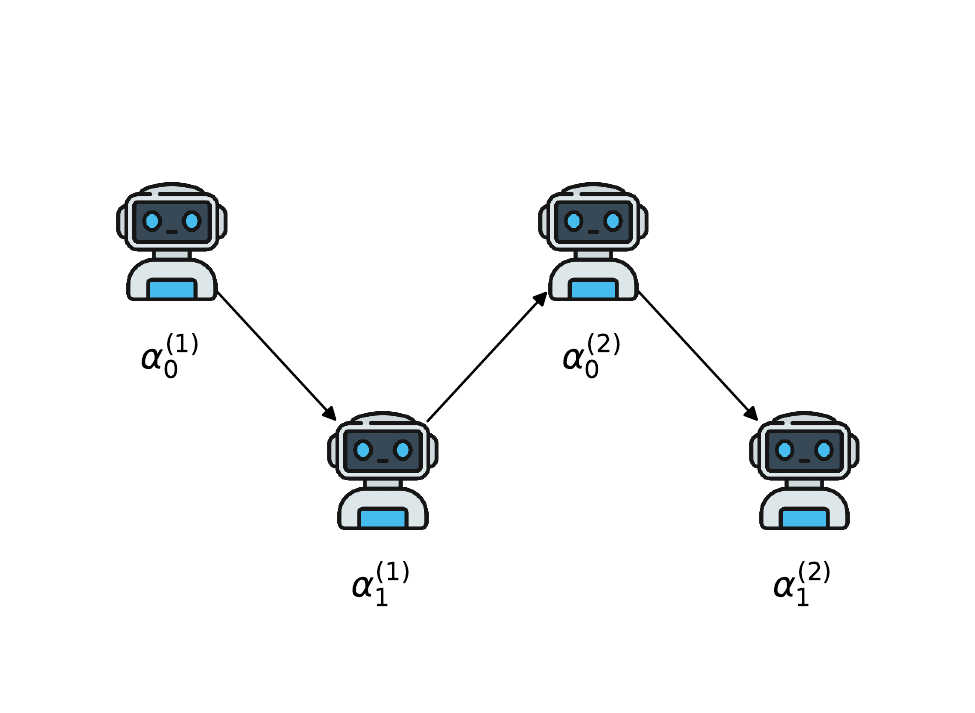}
        \caption{Debate (Db).}
        \label{fig:pattern-debate}
    \end{subfigure}
    \begin{subfigure}[b]{0.33\textwidth}
        \includegraphics[width=\textwidth]{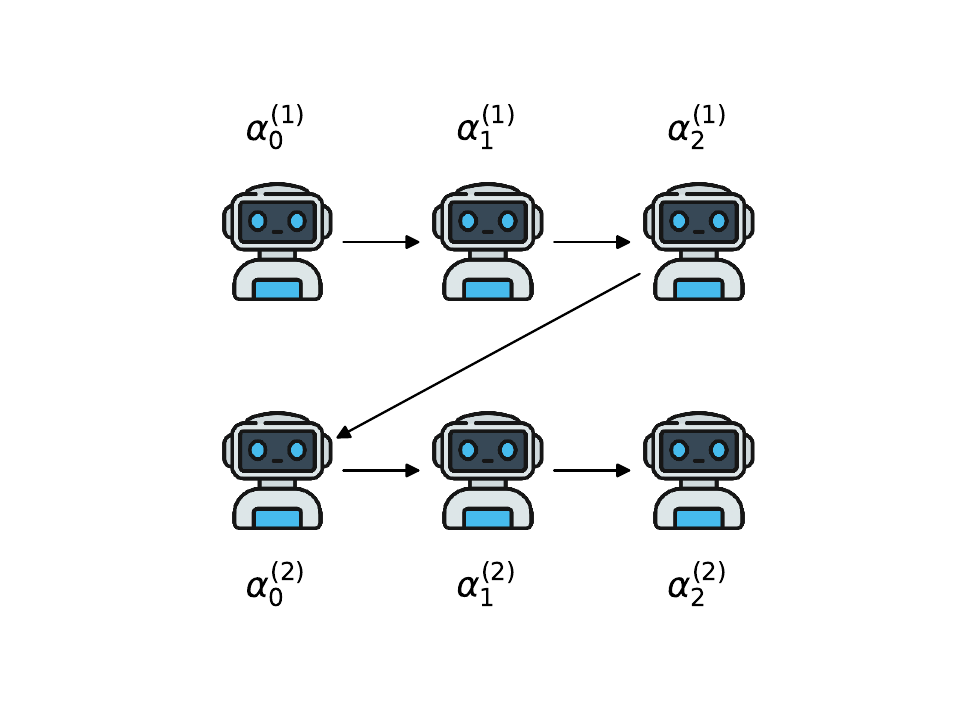}
        \caption{Self-Refine (SR).}
        \label{fig:pattern-refine}
    \end{subfigure}
    \caption{Canonical agentic reasoning patterns: (a) single agent: atomic block; (b) chain-of-thought: sequential chain that refines output agent by agent; (c) self-consistency: parallel agents merged into a consensus agent; (d) tree-of-thought: one agent fanning out into multiple parallel agents; (e) debate: two agents debating back and forth over multiple rounds and (f) self-refine: chain of agents revising output over multiple rounds.}
    \label{fig:patterns}
\end{figure*}

A \textit{single-agent} (SA) pattern is the simplest topology consisting of a single agent $\alpha_0$, serving as the atomic block for more complex pattern compositions. A \textit{chain-of-thought} (CoT) pattern is a sequential chain of $k \geq 2$ agents $\alpha_0, \ldots, \alpha_{k-1}$ with $(\alpha_j, \alpha_{j+1}) \in E$ for $j \in \{ 0, \ldots, k-2 \}$. In CoT, each agent refines the output of the previous agent. A \textit{self-consistency} (SC) pattern is $k \geq 2$ agents $\alpha_0, \ldots, \alpha_{k-1}$ that execute in parallel and all merge into a single agent $\alpha_{k}$, with $(\alpha_j, \alpha_{k}) \in E$ for $j \in \{ 0, \ldots, k-1 \}$. A \textit{tree-of-thought} (ToT) pattern introduces parallel exploration, with a single agent $\alpha_0$ fanning out to $k \geq 2$ parallel agents, with $(\alpha_0, \alpha_j) \in E$ for $j \in \{ 1, \ldots, k \}$. A \textit{debate} (Db) pattern models adversarial multi-turn reasoning between two agents back-and-forth over $r \geq 2$ rounds. To achieve acyclicity, we unroll it into $2r$ distinct agent instances as  $\alpha_A^{(1)}, \alpha_B^{(1)}, \ldots, \alpha_A^{(r)}, \alpha_B^{(r)}$ in sequence. Finally, a \textit{self-refine} (SR) pattern is a chain of $k \geq 2$ agents that revises its own output over $r \geq 2$ rounds, which is similarly unrolled into $r$ copies of the $k$-agent chain. Figure~\ref{fig:patterns} illustrates all these six patterns.

A complete workflow is assembled by drawing multiple pattern instances and connecting them into a single weakly connected DAG where the last agent of the previous pattern is linked to the first agent of the next pattern and additional edges are added between agents at random. The order in which the patterns are sequenced follows a first-order Markov chain with pattern-to-pattern transition probabilities. Therefore, the workflow is not said to be merely a disjoint union of isolated patterns. Each workflow is characterized by properties: size $N = |\mathcal{A}|$ as the total number of agents; depth $d(G)$ as the length of the longest path; and width $w(G)$ as the largest number of agents at the same depth level. The construction of a workflow with given characteristics is feasible only when:
\[
w(G) + d(G) \;\leq\; N \;\leq\; w(G) \cdot \bigl(d(G) + 1\bigr)
\]
where $N$ agents are distributed across $d(G) + 1$ depth levels and each agent in level $\ell > 0$ is wired to at least one predecessor in level $\ell - 1$ and each agent in level $\ell < d(G)$ to at least one successor in level $\ell + 1$ and one level is sized to exactly $w(G)$ agents.

\subsection{Heterogeneous Cloud Model}

A cloud resource is a virtualized and on-demand computation instance to which an agent can be assigned for execution. Each resource $r \in \mathcal{R}$ is characterized by a tuple $r = (v_r, g_r, B_r, p_r)$, where $v_r \in \mathbb{Z}_{\geq 0}$ is the vCPU count, $g_r \in \mathbb{Z}_{\geq 0}$ is the vGPU count, $B_r \in \mathbb{R}^+$ is the network bandwidth (Gbps) and $p_r \in \mathbb{R}^+$ is the hourly on-demand price (USD/hour). For the total compute power of a resource, we unify CPU and GPU capacity into a single scalar via a weighted sum as:
\[
\mathrm{cp}(r) \;=\; w_v \cdot v_r \;+\; w_g \cdot g_r
\]
where $w_v, w_g \in \mathbb{R}^+$ are the weights for the relative contribution of a vCPU and a vGPU to inference. For instance, a CPU-only resource ($g_r = 0$) with $w_v = 1$ reduces $\mathrm{cp}(r)$ to only its vCPU count exactly. From the available resources in the cloud, we compose three environments as CPU-only $\mathcal{R}_{\mathrm{cpu}}$ with $g_r = 0$ for all $r$, GPU-accelerated $\mathcal{R}_{\mathrm{gpu}}$ with $g_r > 0$ for $r$ and CPU+GPU including all resources.

When two dependent agents run on different resources, the output of the first agent must be transferred to the second over the network, subject to the available bandwidth. The bottleneck bandwidth between any two resources $r_i, r_j \in \mathcal{R}$ is computed as:
\[
B(r_i, r_j) \;=\; \min(B_{r_i}, B_{r_j})
\]
where $B(r_i, r_j)$ is the bandwidth between $r_i$ and $r_j$. No network transfer occurs when two agents are hosted exactly on the same resource. In this work, we consider $|\mathcal{R}_{\mathrm{cpu}}| = 30$ AWS EC2 CPU-only instances (see Table~\ref{tab:cloud_resources}) and $|\mathcal{R}_{\mathrm{gpu}}| = 30$ AWS EC2 GPU-accelerated instances (see Table~\ref{tab:gpu_resources}). This resource catalog serves solely to parameterize $v_r$, $g_r$, $B_r$ and $p_r$ with concrete values for our simulation and our cloud model itself remains independent of this particular set of resources. Notably, our unified computation model treats CPU and GPU capacity as interchangeable, which is very straightforward. This might result in assigning large models to CPU-only resources. Although such assignments are impractical in real executions, they are admissible under our model which does not intend to capture real executions. For the sake of simplicity, no workflow-driven or cloud-driven uncertainty (e.g., fluctuation in inference latency, network congestion) is considered in this work.

\subsubsection{CPU-Only Environment $\mathcal{R} \subseteq \mathcal{R}_{\mathrm{cpu}}$} This environment takes into account only the resources with no active GPU $g_r = 0$. This deliberate limitation on resource heterogeneity naturally results in certain scheduling impacts. First, unlike the GPU-only environment, there is no hardware-driven improvement on latency and the schedulers cannot buy more speed for the critical-path agents and must instead rely more heavily on parallelizing agents across many resources. Second, per-hour prices of CPU-only resources are cheaper when compared to the resources with GPU, resulting in lower total resource costs at the end of the optimization. In addition, CPU-only resources span a finer-grained range of vCPU counts and prices, which gives the schedulers a more fine-grained makespan--cost trade-off. In Table~\ref{tab:cloud_resources}, the specifications of the CPU-only instance families (M1, M3, M4, M5, M5D) considered are listed, spanning vCPU counts from 1 to 96, network bandwidth from 10 to 25 Gbps and hourly pricing from \$0.044 to \$5.424.

\subsubsection{GPU-Accelerated Environment $\mathcal{R} \subseteq \mathcal{R}_{\mathrm{gpu}}$} Every resource in this environment is GPU-accelerated $g_r > 0$, offering more computational power per unit than a CPU-only environment. Here, agents complete their inference faster, compressing makespan toward the lower values. However, this compression is not free since GPU-accelerated resources incur considerably higher per-hour costs in return. Therefore, this environment challenges the schedulers to still find a meaningful cost--utilization trade-off once the makespan objective is largely satisfied by hardware alone. In Table~\ref{tab:gpu_resources}, the specifications of the GPU-accelerated instance families (G4dn, G5, G6, G6e, G7, G7e, P4d, P4de, P5 and P5en) are listed, spanning several NVIDIA GPU generations, from the T4 through the H200.

\subsubsection{CPU+GPU Environment $\mathcal{R} \subseteq \mathcal{R}_{\mathrm{cpu}} \cup \mathcal{R}_{\mathrm{gpu}}$} This environment considers both CPU-only and GPU-accelerated resources at the same time for agent assignment. This provides a broader scheduling since organizations generally provision elastic cloud capacity in practice and relax the schedulers to make workload-specific decisions about which agent runs on which resource. In return, this hardens the overall optimization since the schedulers now have to map an agent on a cheaper but moderate-throughput CPU resource or a more expensive but high-throughput GPU resource or their hybrid composition. For instance, an agent on the critical path might benefit more from a GPU to reduce latency for the agents immediately waiting on it. Consequently, achievable Pareto-fronts for this environment are likely to be more diverse, spanning the full range from all-CPU to all-GPU scheduling solutions and every combination in between.

\subsection{Workflow Simulation Model in Cloud}
\label{sec:simulation}

Given a resource assignment $\sigma: \mathcal{A} \to \mathcal{R}$, the simulation model executes that assignment by considering the characteristics of workflows and cloud computing to derive makespan, cost and utilization. In this paper, we assume a non-preemptive execution where an agent runs to completion without interruption once it starts on a resource. An agent must jointly satisfy two conditions to start: \textit{dependency readiness} where all of its predecessors must finish and transfer their data and \textit{resource availability} where the resource must be free of any execution. For every agent $\alpha$, its ready time $\mathrm{RT}[\alpha]$ is the earliest moment the \textit{dependency readiness} is satisfied:
\[
\mathrm{RT}[\alpha] \;=\; \max_{\alpha' \in \mathrm{pred}(\alpha)} \Bigl( \mathrm{FT}[\alpha'] + \tilde{o}(\alpha')/B(\sigma(\alpha'), \sigma(\alpha)) \Bigr)
\]
where $\mathrm{FT}$ is the finish time and $\mathrm{RT}$ is zero if $\alpha$ has no predecessors and the transfer term is zero when $\alpha'$ and $\alpha$ share the same resource. Here, $\tilde{o}(\alpha')$ is the data size of the output of $\alpha'$, obtained as $\tilde{o}(\alpha') = 4 \cdot o(\alpha')$ bytes. Here, we assume that one token corresponds to four single-byte characters of English text.

\begin{figure*}[htbp]
\centerline{\includegraphics[width=0.72\textwidth]{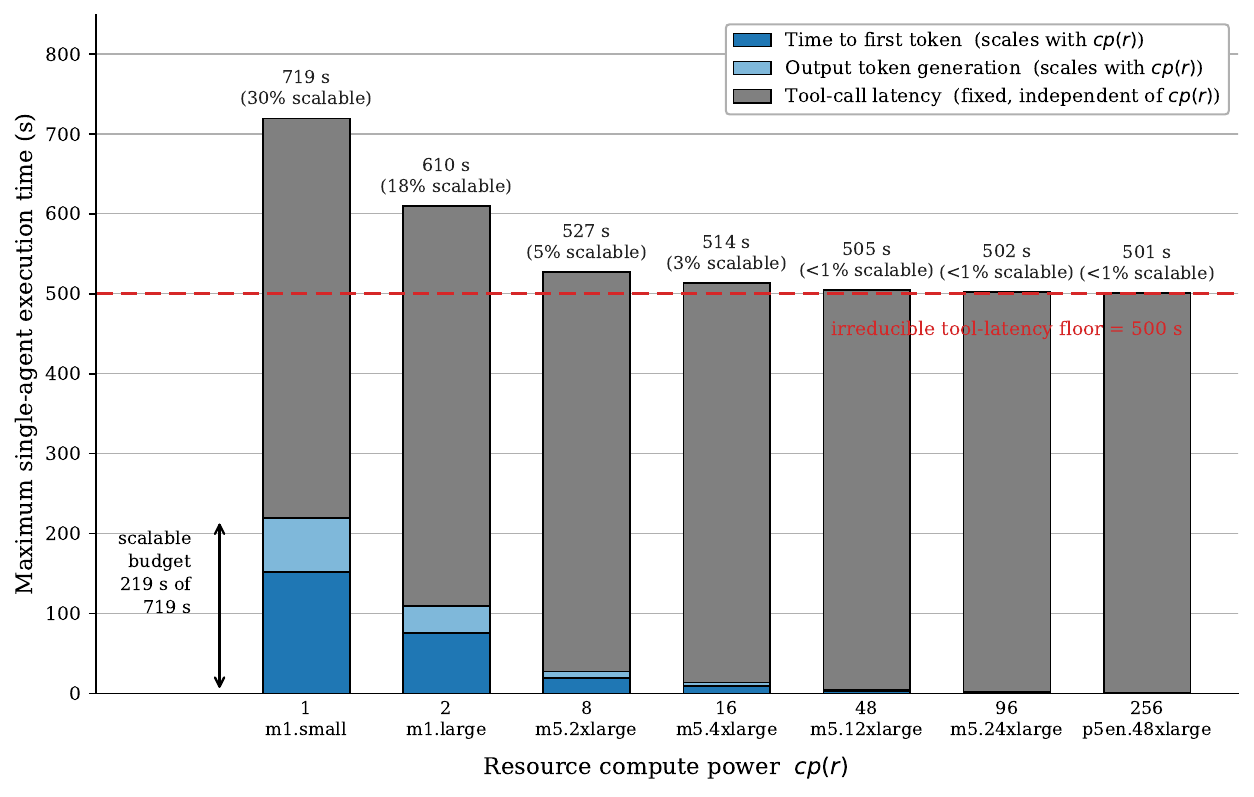}}
\caption{Single-agent execution time against resource computation power $\mathrm{cp}(r)$. The TTFT and the output-token generation time scale with $\mathrm{cp}(r)$ while the total tool-call latency does not. Even with the highest catalog TTFT, 5000 output tokens and five tools issuing ten calls of ten seconds each, only about $30\%$ of the budget is scalable.}
\label{fig:makespan_scalability}
\end{figure*}

Resource availability needs some exclusive cases to be properly managed. If the resource $r$ has never been used before, it must first be provisioned with a Gaussian-distributed initialization delay $\mathrm{InitTime}(r)$ before $\alpha$ can start. If $r$ has been used before but its billing period is closed, it must be re-provisioned exactly as if it were new, by incurring $\mathrm{InitTime}(r)$ again. Lastly, if $r$ is still actively provisioned and free, $r$ can start a new agent task with no additional delay. Once the start time $\mathrm{ST}[\alpha]$ of $\alpha$ is set with these cases, its execution time can be computed as:
\[
\mathrm{ET}[\alpha] \;=\; L_{\mathrm{inf}}(\alpha)/\mathrm{cp}(r) + L_{\mathrm{tool}}(\alpha)
\]
where the inference time is scaled by the resource computation power. Its finish time is $\mathrm{FT}[\alpha] = \mathrm{ST}[\alpha] + \mathrm{ET}[\alpha]$. Figure~\ref{fig:makespan_scalability} shows this decomposition for the maximum single-agent budget. Only the inference term $L_{\mathrm{inf}}(\alpha)/\mathrm{cp}(r)$ depends on the resource an agent runs on. Even for the highest TTFT in the model catalog together with the maximum output length, only about $30\%$ of the budget at $\mathrm{cp}(r)=1$ is scalable by $\mathrm{cp}(r)$. On the other hand, the tool term $L_{\mathrm{tool}}(\alpha)$ is always a fixed floor that no resource can reduce. A scheduler can therefore compress the execution time of an individual agent only within a narrow band.

The billing period of a resource starts when it is provisioned and closes whenever it is de-provisioned. Note that de-provisioning also consumes a Gaussian-distributed termination delay as $\mathrm{TermTime}(r)$ before the period actually closes. Every billing period is charged in full hours, rounded up with an hourly on-demand price. The workflow completes when all exit agents $\mathcal{A}_{\mathrm{exit}} = \{\alpha \in \mathcal{A} : \mathrm{succ}(\alpha) = \emptyset\}$ have finished their executions. 

\begin{algorithm*}
\caption{Workflow Simulation in Cloud.}
\label{alg:simulation}
\begin{algorithmic}[1]
\Require Workflow DAG $G = (\mathcal{A}, E)$; resource assignment $\sigma: \mathcal{A} \to \mathcal{R}$
\Ensure Schedule; makespan $M$; cost $C$; utilization $U$
\State $\mathrm{order} \gets$ topological order of $G$ (ties broken uniformly at random)
\State $\mathrm{FT}[\cdot], \mathrm{lastFT}[\cdot], \mathrm{billingStart}[\cdot], \mathrm{billing}[\cdot] \gets \emptyset$
\For{$\alpha \in \mathrm{order}$}
    \State $r \gets \sigma(\alpha)$
    \State $\mathrm{RT}[\alpha] \gets \max_{\alpha' \in \mathrm{pred}(\alpha)} \bigl( \mathrm{FT}[\alpha'] + \tilde{o}(\alpha') / B(\sigma(\alpha'), r) \bigr)$ \Comment{$0$ if $\mathrm{pred}(\alpha) = \emptyset$}
    \If{$r$ not yet used}
        \State $\mathrm{ST}[\alpha] \gets \mathrm{RT}[\alpha] + \mathrm{InitTime}(r)$ \Comment{Gaussian provisioning delay}
        \State $\mathrm{billingStart}[r] \gets \mathrm{RT}[\alpha]$
    \ElsIf{$\mathrm{lastFT}[r] \leq \mathrm{RT}[\alpha]$} \Comment{$r$ has gone idle: close its billing period, re-provision}
        \State $\mathrm{billing}[r] \gets \mathrm{billing}[r] \cup \{(\mathrm{billingStart}[r],\ \mathrm{lastFT}[r] + \mathrm{TermTime}(r))\}$
        \State $\mathrm{ST}[\alpha] \gets \mathrm{RT}[\alpha] + \mathrm{InitTime}(r)$
        \State $\mathrm{billingStart}[r] \gets \mathrm{RT}[\alpha]$
    \Else \Comment{$r$ is still running from a previous agent}
        \State $\mathrm{ST}[\alpha] \gets \mathrm{lastFT}[r]$
    \EndIf
    \State $\mathrm{ET}[\alpha] \gets L_{\mathrm{inf}}(\alpha)/\mathrm{cp}(r) + L_{\mathrm{tool}}(\alpha)$
    \State $\mathrm{FT}[\alpha] \gets \mathrm{ST}[\alpha] + \mathrm{ET}[\alpha]$
    \State $\mathrm{lastFT}[r] \gets \mathrm{FT}[\alpha]$
\EndFor
\For{$r$ with an open billing period}
    \State $\mathrm{billing}[r] \gets \mathrm{billing}[r] \cup \{(\mathrm{billingStart}[r],\ \mathrm{lastFT}[r] + \mathrm{TermTime}(r))\}$ \Comment{close out}
\EndFor
\State $M \gets \max_{\alpha \in \mathcal{A}} \mathrm{FT}[\alpha]$
\State $C \gets \sum_{r \in \mathcal{R}} \sum_{(t_s, t_e) \in \mathrm{billing}[r]} \lceil (t_e - t_s)/3600 \rceil \cdot p_r$
\State $U \gets \dfrac{1}{|R(\sigma)|} \displaystyle\sum_{r \in R(\sigma)} \dfrac{\sum_{\alpha \,:\, \sigma(\alpha) = r} \mathrm{ET}[\alpha]}{\sum_{(t_s,t_e) \in \mathrm{billing}[r]} (t_e - t_s)}$ \Comment{busy time over billed time, averaged over the resources $\sigma$ actually uses}
\State \Return schedule, $M$, $C$, $U$
\end{algorithmic}
\end{algorithm*}

\subsection{Multi-Agent Workflow Scheduling as Multi-Objective Optimization}
\label{sec:problem-optimization}

A multi-objective optimization problem (MOP) is a problem with two or more objectives where at least two objectives are in conflict. Without loss of generality, a MOP over a decision space $X$ can be formally defined in the following way:
\begin{align*}
\min_{x \in X} & \quad F(x) \;=\; \bigl(f_1(x), \ldots, f_k(x)\bigr)^T, \\
\text{s.t.} & \quad g_i(x) \leq 0,\ i = 1, \ldots, p, \\
& \quad h_j(x) = 0,\ j = 1, \ldots, q,
\end{align*}
where $F: X \to \mathbb{R}^k$, $k \geq 2$, maps each solution $x$ in decision space into a point in objective space $\mathbb{R}^k$ under the inequality and equality constraint functions $g_i$ and $h_j$. Because of the conflicting objectives (i.e., improving an objective comes at the expense of another), there is typically no single $x^* \in X$ that simultaneously optimizes every $f_i$. Therefore, a MOP represents such trade-offs via the concept of Pareto-domination. For a minimization problem, a solution $x$ Pareto-dominates $x'$ ($x \prec x'$) if $f_i(x) \leq f_i(x')$ for every $i \in \{1, \ldots, k\}$ and $f_i(x) < f_i(x')$ for at least one $i$. In other words, $x$ is at least as good as $x'$ on every objective and strictly better on at least one. A solution $x^* \in X$ is Pareto-optimal if no solution in $X$ dominates it, $\nexists\, x \in X : x \prec x^*$ where the set of every such solution is the Pareto-optimal set (POS) in decision space and its projection under $F$ in objective space is the Pareto-optimal front (POF). 

\subsubsection{Decision Variables} The decision space for our agentic workflow scheduling problem is the set of all the resource assignment solutions as:
\[
X \;=\; \{\, \sigma : \mathcal{A} \to \mathcal{R} \,\}
\]
where the function maps every agent in the workflow to exactly one resource in the cloud. The schedulers search over the entire decision space to seek such a feasible assignment solution $\sigma$. For the sake of simplicity, we assume that each resource can execute at most one agent at a time, without considering the resource-level parallelism at the moment.

\subsubsection{Objective Functions} Every $\sigma \in X$ is evaluated by simulating its execution (see Algorithm~\ref{alg:simulation}) and collecting the resulting makespan, cost and utilization values. This lets us state the multi-agent workflow scheduling problem as the multi-objective optimization problem:
\[
\min_{\sigma \in X} \quad F(\sigma) = (M(\sigma), C(\sigma), -U(\sigma))^T: X \to \mathbb{R}^3
\]
where the makespan objective $M(\sigma)$ is the completion time of the last agent to finish: 
\[
M(\sigma) \;=\; \max_{\alpha \in \mathcal{A}} \mathrm{FT}[\alpha]
\]
$\mathrm{FT}[\alpha]$ is the finish time of agent task $\alpha$. The cost objective is the total monetary cost across the billing periods of all resources:
\[
C(\sigma) \;=\; \sum_{r \in \mathcal{R}} \sum_{(t_s, t_e) \in \mathrm{billing}[r]} \Bigl\lceil \frac{t_e - t_s}{3600} \Bigr\rceil \cdot p_r
\]
where each billing period is rounded up to the nearest full hour and charged at its hourly on-demand price $p_r$. Note that the inner sum counts the number of cycles the resource $r$ is provisioned and de-provisioned. The utilization objective refers to the average resource busy time as:
\[
U(\sigma) \;=\; \frac{1}{|R(\sigma)|} \sum_{r \in R(\sigma)} \frac{\sum_{\alpha \,:\, \sigma(\alpha) = r} \mathrm{ET}[\alpha]}{\sum_{(t_s,t_e) \in \mathrm{billing}[r]} (t_e - t_s)}
\]
where $\mathrm{ET}[\alpha]$ is the execution time of agent $\alpha$ and $R(\sigma)$ is the set of resources to which $\sigma$ assigns at least one agent. In this regard, $M(\sigma)$ and $C(\sigma)$ are in conflict since faster resources incur higher on-demand prices. $U(\sigma)$ is also in conflict with $M(\sigma)$ since maximizing busy times favors using fewer and heavily-loaded resources, which lowers parallelism.

\subsubsection{Optimization Constraints} The feasible region of a MOP is limited through the inequality and equality constraints $g_i(x) \leq 0$ and $h_j(x) = 0$. Our agentic workflow scheduling problem involves four such constraints. The \emph{precedence constraint} asks that no agent begin execution before all its predecessor agents have finished and all data has fully arrived:
\[
\mathrm{ST}[\alpha] \;\geq\; \mathrm{FT}[\alpha'] + \tilde{o}(\alpha')/B(\sigma(\alpha'), \sigma(\alpha)), \qquad \forall\, (\alpha', \alpha) \in E
\]
The \emph{resource serialization constraint} requires that no resource execute two agents concurrently. For any two distinct agents $\alpha, \alpha' \in \mathcal{A}$ sharing a resource, $\sigma(\alpha) = \sigma(\alpha') = r$, their execution intervals should not overlap:
\[
\mathrm{ST}[\alpha] \geq \mathrm{FT}[\alpha'] \qquad \text{or} \qquad \mathrm{ST}[\alpha'] \geq \mathrm{FT}[\alpha]
\]
The \emph{assignment completeness constraint} requires that every agent be assigned to exactly one resource:
\[
\forall\, \alpha \in \mathcal{A},\ \exists!\, r \in \mathcal{R} : \sigma(\alpha) = r
\]
and finally the \emph{non-negativity constraint} requires that no agent start before the workflow itself begins:
\[
\mathrm{ST}[\alpha] \geq 0, \quad \forall \alpha \in \mathcal{A}
\]
The schedulers in MO-SAIC are designed to satisfy all these optimization constraints by construction, so no explicit constraint check or repair mechanism is necessary during solution construction. 

\subsubsection{Pareto-Optimality} 
\begin{figure}[htbp]
\centerline{\includegraphics[width=0.50\textwidth]{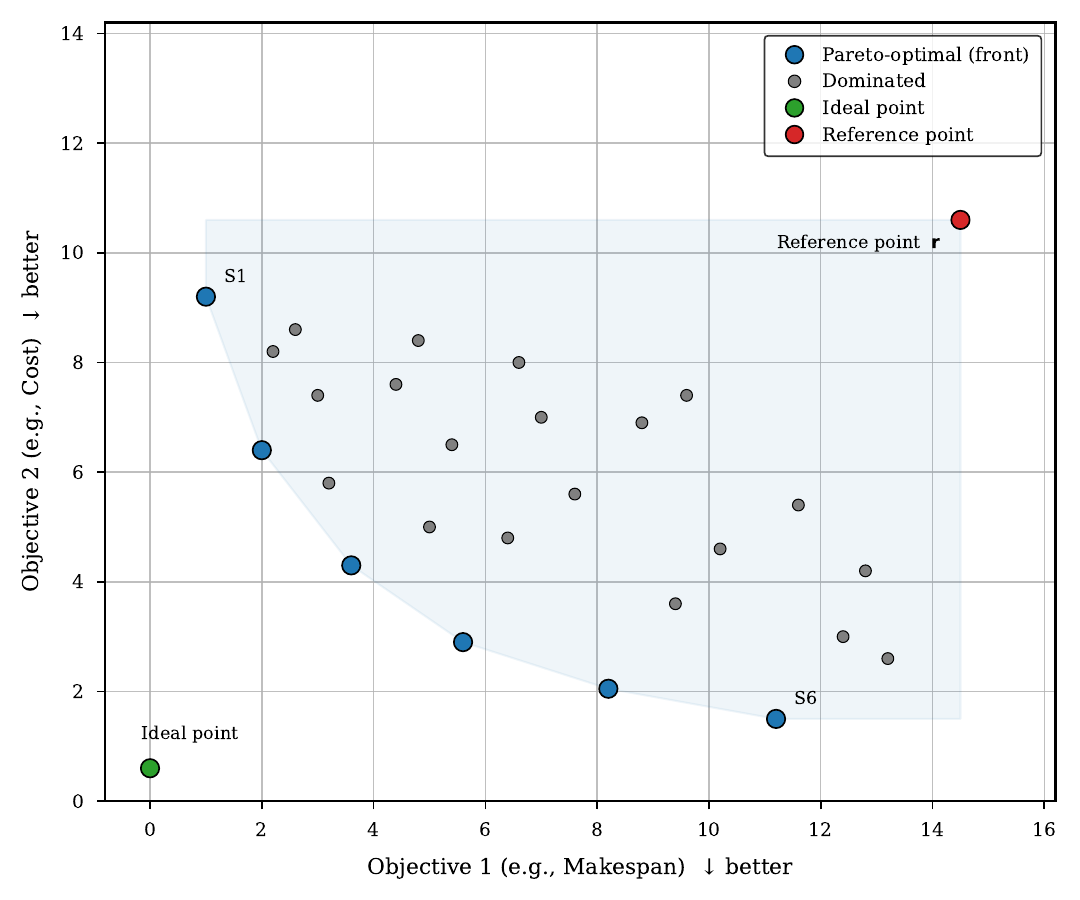}}
\caption{An illustrative Pareto front over two minimization objectives with non-dominated solutions (in blue) and dominated solutions (in gray). The shading shows the region both objectives are simultaneously worse than the front solutions. The ideal point is the best-possible solution while the reference point is the anchor for hypervolume calculations \cite{zitzler1998}.}
\label{fig:pareto_illustration}
\end{figure}

A scheduling solution $\sigma$ is said to weakly dominate $\sigma'$ ($\sigma \preceq \sigma'$) if:
\[
\begin{aligned}
M(\sigma) &\leq M(\sigma') \\
C(\sigma) &\leq C(\sigma') \\
U(\sigma) &\geq U(\sigma')
\end{aligned}
\]
and $\sigma$ Pareto-dominates $\sigma'$ ($\sigma \prec \sigma'$) if additionally at least one of these three holds strictly. A schedule $\sigma^* \in X$ is Pareto-optimal if no $\sigma \in X$ dominates it where the resulting POS is:
\[
X^* \;=\; \{\, \sigma^* \in X : \nexists\, \sigma \in X,\ \sigma \prec \sigma^* \,\}
\]
and its projection $F(X^*)$ in objective space is the POF. However, the decision space of the problem we define is far too large to search exhaustively in the complexity level of $|\mathcal{R}|^{|\mathcal{A}|}$. In other words, $X^*$ is intractable to compute exactly. This is why we should employ a strong metaheuristic to solve the problem or, more precisely, to approximate the non-dominated front. Figure~\ref{fig:pareto_illustration} illustrates Pareto-domination in two-dimensional objective space by distinguishing non-dominated solutions from the dominated solutions.

\section{MO-SAIC Framework}
\label{sec:framework}

This section presents MO-SAIC, our multi-objective multi-agent workflow scheduling framework. Every scheduler in MO-SAIC searches over the same decision space $X$ and is evaluated through the same workflow simulation environment against the same three objectives of makespan, cost and utilization. The schedulers differ only in how they explore $X$ without any control over how their solutions are evaluated, so their results are comparable. 

\subsection{Framework Architecture}

MO-SAIC is composed of five components: (i) workflow generator, generating individual agentic workflows with given size, depth and width parameters; (ii) cloud, providing the heterogeneous CPU and GPU resource pool to which the agents can be assigned; (iii) schedulers, searching the resulting decision space for the optimal scheduling solutions; (iv) simulator, executing candidate scheduling solutions to reveal the objective vector; and (v) evaluator, ranking the population of scheduling solutions by non-dominated sorting and crowding distance to reveal the Pareto-optimal front. For the fair comparison of the schedulers, we keep all other components identical across the framework. As a pipeline, the workflow generator first generates a single workflow instance. Secondly, the schedulers take this instance and the available cloud resources to generate a population of scheduling solutions. Thirdly, regardless of which scheduler produced them, the simulator evaluates all solutions in the same way. Lastly, evaluator ranks the solutions to feed back to the schedulers until the maximum generation $\Gamma_{\max}$ is reached and later to select the Pareto-optimal solutions.

\subsection{NSGA-II Scheduler}

NSGA-II~\cite{deb2002} is an elitist multi-objective genetic algorithm that evolves a population of solutions across generations by using: (i) non-dominated sorting to rank solutions via Pareto-domination into several ranks; and (ii) crowding distance to preserve diversity in the same ranks. In this subsection, we customize NSGA-II to work with the solution representation and optimization objectives of our multi-objective agentic workflow scheduling problem. Refer to Algorithm~\ref{alg:nsga2} for the complete algorithm.

\subsubsection{Solution Representation} Each solution chromosome $\sigma \in X$ is represented as a string of decision variables with the length of the number of available agents in workflow $|\mathcal{A}|$. Each gene in a chromosome encodes the resource to which the agent on that position is assigned. During the execution of that agent (see Section~\ref{sec:simulation}), each gene is decoded to assign the agent to its corresponding resource.

\subsubsection{Population Initialization} The initial population of NSGA-II $\mathcal{P}_0 = \{\sigma_0, \ldots, \sigma_{\lambda-1}\}$ is drawn by assigning every agent to an independent and uniformly random resource as:
\[
\sigma_i(\alpha) \sim \mathrm{Uniform}(\mathcal{R}), \qquad \forall\, \alpha \in \mathcal{A},\ i = 0, \ldots, {\lambda-1}
\]
where $\lambda$ is the number of solutions in the population. All the schedulers in this framework are randomly initialized, which implies they all start from statistically indistinguishable initial populations. Therefore, every improvement reported in the experimental study can be directly attributed to the search space exploration capabilities of the schedulers themselves. 

\subsubsection{Fitness Evaluation} Every chromosome is evaluated at every generation with respect to the makespan, cost and utilization objectives of the simulation in cloud. The resulting objective vector $F(\sigma) = (M(\sigma), C(\sigma), -U(\sigma))$ is preserved alongside the solution for the rest of its lifetime across the generations. Refer to Section~\ref{sec:problem-optimization} for the calculations of these objectives.

\subsubsection{Non-Dominated Sorting} Given the population $\mathcal{P}$ with $2\lambda$ individuals during environmental selection, the non-dominated sorting partitions it into successive fronts $F_0(\mathcal{P}), F_1(\mathcal{P}), \ldots$ by using Pareto-domination to select the best $\lambda$ solutions. $F_0(\mathcal{P})$ represents the Pareto-optimal front, having the solutions $\sigma \in \mathcal{P}$ that are not dominated by any other solution in $\mathcal{P}$. The next front $F_1(\mathcal{P})$ has every remaining solution that is not dominated by any other solution where it goes until the entire population is exhausted.

\subsubsection{Crowding Distance} Within a single front $F_k(\mathcal{P})$, the crowding distance $\mathrm{cd}(\sigma)$ calculates how isolated $\sigma$ is from its neighbors in the same front to push evolution to sparsely-explored regions. For each objective, the front is first sorted and the two solutions sitting at the two ends of that objective are assigned $\mathrm{cd} = \infty$ to always preserve the front boundaries. Every interior solution $\sigma$ accumulates its value as:
\[
\mathrm{cd}(\sigma) \mathrel{+}= \frac{f(\sigma_{i+1}) - f(\sigma_{i-1})}{f_{\max} - f_{\min}}
\]
where $f_{\min}, f_{\max}$ are the range of that objective and $\mathrm{cd}(\sigma)$ are summed over all objectives. A solution in a more empty neighborhood has a larger $\mathrm{cd}(\sigma)$.

\subsubsection{Parent Selection} Parents are drawn via tournament selection where a certain number of chromosomes are first randomly sampled from the population and the best chromosome with respect to the rank and crowding distance wins the tournament. Every child needs two independently drawn parents from two independent tournaments. Pressure through rank and crowding distance naturally pushes parent selection toward a better Pareto-optimal front with less-crowded regions. 

\subsubsection{Crossover and Mutation} Once two parents $\sigma_a, \sigma_b$ are selected, the two-point crossover is applied by partitioning each chromosome into three contiguous segments. Then, the child inherits the outer two segments from $\sigma_a$ and the middle segment from $\sigma_b$. Every gene of the resulting child is then independently mutated with the pre-specified mutation rate (see Table~\ref{table:parameters}), which reassigns that agent to a new uniformly random resource in $\mathcal{R}$.

\subsubsection{Elitism} NSGA-II maintains the elitism (i.e., preserving the best solutions ever found across generations) by default. Once $F_0(\mathcal{P})$ is found, the population for the next generation is filled from $F_0(\mathcal{P})$ first before any lower-ranked front is even considered. Hence, no non-dominated solution the population has ever found is discarded by a worse solution.

\begin{algorithm*}
\caption{NSGA-II Scheduler.}
\label{alg:nsga2}
\begin{algorithmic}[1]
\Require Agent set $\mathcal{A}$; resource set $\mathcal{R}$; population size $\lambda$; generation count $\Gamma_{\max}$; mutation rate $p_m$; tournament size $\tau$
\Ensure Final population $\mathcal{P}$ of $\lambda$ resource assignments
\State $\mathcal{P} \gets \{\sigma_0, \ldots, \sigma_{\lambda-1}\}$, each $\sigma_i(\alpha) \sim \mathrm{Uniform}(\mathcal{R})\ \forall\, \alpha \in \mathcal{A}$ \Comment{population initialization}
\State $\mathrm{Sim} \gets \{\mathrm{Simulate}(\sigma) : \sigma \in \mathcal{P}\}$ \Comment{Algorithm~\ref{alg:simulation}, cached per individual}
\For{$\mathrm{gen} = 1$ to $\Gamma_{\max}$}
    \State $F_0(\mathcal{P}), F_1(\mathcal{P}), \ldots \gets \mathrm{NonDominatedSort}(\mathrm{Sim})$
    \State $\mathrm{cd}(\cdot) \gets \mathrm{CrowdingDistance}\bigl(F_0(\mathcal{P}), F_1(\mathcal{P}), \ldots\bigr)$
    \State $\mathrm{rank}(\sigma) \gets k$ such that $\sigma \in F_k(\mathcal{P})$, for every $\sigma \in \mathcal{P}$
    \State $\mathrm{Children} \gets \emptyset$
    \While{$|\mathrm{Children}| < \lambda$}
        \For{$i \in \{a, b\}$} \Comment{two independently drawn parents}
            \State $S \gets \mathrm{TournamentSelect}( \mathcal{P}$)
            \State $\mathrm{parent}_i \gets \arg\min_{\sigma \in S} \bigl( \mathrm{rank}(\sigma),\, -\mathrm{cd}(\sigma) \bigr)$ \Comment{crowded-comparison}
        \EndFor
        \State $\sigma_{\mathrm{child}} \gets \mathrm{Crossover}(\mathrm{parent}_a, \mathrm{parent}_b)$ \Comment{two-point over fixed agent order}
        \State $\mathrm{Mutate}(\sigma_{\mathrm{child}}, p_m)$ \Comment{per-gene reassignment}
        \State $\mathrm{Children} \gets \mathrm{Children} \cup \{\sigma_{\mathrm{child}}\}$
    \EndWhile
    \State $\mathrm{Sim}_{\mathrm{children}} \gets \{\mathrm{Simulate}(\sigma) : \sigma \in \mathrm{Children}\}$
    \State $\mathrm{Pool} \gets \mathcal{P} \cup \mathrm{Children}$
    \State $\mathrm{PoolResults} \gets \mathrm{Sim} \cup \mathrm{Sim}_{\mathrm{children}}$
    \State $\mathcal{P} \gets \mathrm{SelectBest}(\mathrm{Pool}, \mathrm{PoolResults}, \lambda)$ \Comment{elitist replacement}

\EndFor
\State \Return $\mathcal{P}$
\end{algorithmic}
\end{algorithm*}

\subsection{Weighted-Sum Genetic Algorithm Scheduler}

W-GA \cite{murata1995}, shares the same solution representation, population initialization, crossover and mutation operators with NSGA-II. The difference from NSGA-II is that W-GA replaces the non-dominated sorting and crowding distance operators with a single scalar fitness score operator with respect to a dynamic weight vector of $w(\sigma) = (w_M, w_C, w_U)$. At every generation, every individual $\sigma$ is assigned its own independently drawn random weight vector $w(\sigma)$ with $w_M + w_C + w_U =1$. In other words, no two individuals and no two generations are scored against the same weights, which pushes the population toward different regions of search space for more front diversity. The fitness score function is defined in the following way:
\[
w_M \cdot \frac{M(\sigma) - M_{\min}}{M_{\max} - M_{\min}} \;+\; w_C \cdot \frac{C(\sigma) - C_{\min}}{C_{\max} - C_{\min}} \;+\; w_U \cdot \left( 1 \;-\; \frac{U(\sigma) - U_{\min}}{U_{\max} - U_{\min}} \right)
\]
where $M_{\min}, M_{\max}$ are the smallest and largest makespan in a generation; and $C_{\min}, C_{\max}$ and $U_{\min}, U_{\max}$ are defined analogously for cost and utilization, respectively. To maintain elitism across generations, two individuals with the best fitness scores are moved unmodified into the next generation while the remainder of the next population is filled through the tournament selection, crossover and mutation operators, as in NSGA-II.

\subsection{MO-HEFT Scheduler}

MO-HEFT \cite{durillo2012} is a multi-objective adaptation of the classical HEFT heuristic \cite{topcuoglu1999}. In this technique, the mean execution time of each agent over all resources in $\mathcal{R}$ is computed as:
\[
\overline{L}(\alpha) \;=\; \frac{1}{|\mathcal{R}|} \sum_{r \in \mathcal{R}} \left( \frac{L_{\mathrm{inf}}(\alpha)}{\mathrm{cp}(r)} + L_{\mathrm{tool}}(\alpha) \right)
\]
where $\overline{L}(\alpha)$ refers to average execution time. The average transfer time as averaged over every ordered pair of resources is:
\[
\overline{T}(\alpha) \;=\; \frac{1}{|\mathcal{R}|^2} \sum_{r_i \in \mathcal{R}} \sum_{r_j \in \mathcal{R}} \frac{\tilde{o}(\alpha)}{B(r_i, r_j)}
\]
where $\overline{T}(\alpha)$ refers to average transfer time. The upward rank is then defined recursively backward:
\[
\mathrm{rank}_u(\alpha) \;=\; \overline{L}(\alpha) \;+\; \begin{cases} 0, & \mathrm{succ}(\alpha) = \emptyset \\ \overline{T}(\alpha) + \max_{\alpha' \in \mathrm{succ}(\alpha)} \mathrm{rank}_u(\alpha'), & \text{otherwise} \end{cases}
\]
where $\mathrm{rank}_u(\alpha)$ is the rank of agent $\alpha$. Once all ranks are computed, agents are ordered by decreasing $\mathrm{rank}_u$ with always a valid topological order. Before the actual assignment, a single baseline solution $\sigma_{\mathrm{base}}: \mathcal{A} \to \mathcal{R}$ is computed with a single-objective HEFT for the minimization of makespan. This baseline solution is never part of the population to be measured. MO-HEFT then processes agents one at a time in priority order by maintaining a front of $\lambda$ partial assignments. It extends every partial assignment for each agent $\alpha$ on the front with every resource choice in $\mathcal{R}$, eventually producing $\lambda|\mathcal{R}|$ candidate solutions. These $\lambda|\mathcal{R}|$ candidates are then reduced back down to $\lambda$ solutions with the non-dominated sorting and crowding distance operators, as in NSGA-II. Once every agent is visited, the partial assignments turn into the full assignments $\sigma \in X$, which MO-HEFT returns as its final population.

\subsection{Random Scheduler}

We include the random scheduler in the framework as the baseline algorithm to evaluate the algorithmic value of the other schedulers. Given the set of agents $\mathcal{A}$ in the workflow and the set of resources in the cloud $\mathcal{R}$, each solution is a random mapping $\sigma: \mathcal{A} \to \mathcal{R}$ with drawing uniformly-random resource for every agent:
\[
\sigma(\alpha) \sim \mathrm{Uniform}(\mathcal{R}), \qquad \forall\, \alpha \in \mathcal{A}
\]
where the scheduler returns a population $\mathcal{P} = \{\sigma_0, \ldots, \sigma_{\lambda-1}\}$ and $\lambda$ is the number of solutions in the population. All four schedulers are compared on the same number of solutions $\lambda$. The random scheduler also follows the problem constraints given in Section~\ref{sec:problem}. Hence, before the resource assignment, it first performs dependency-aware ordering through a topological sort where agents at the same dependency level are selected and randomly shuffled. 

\begin{figure*}[htbp]
\centering
\begin{subfigure}[b]{0.33\textwidth}
    \includegraphics[width=\textwidth]{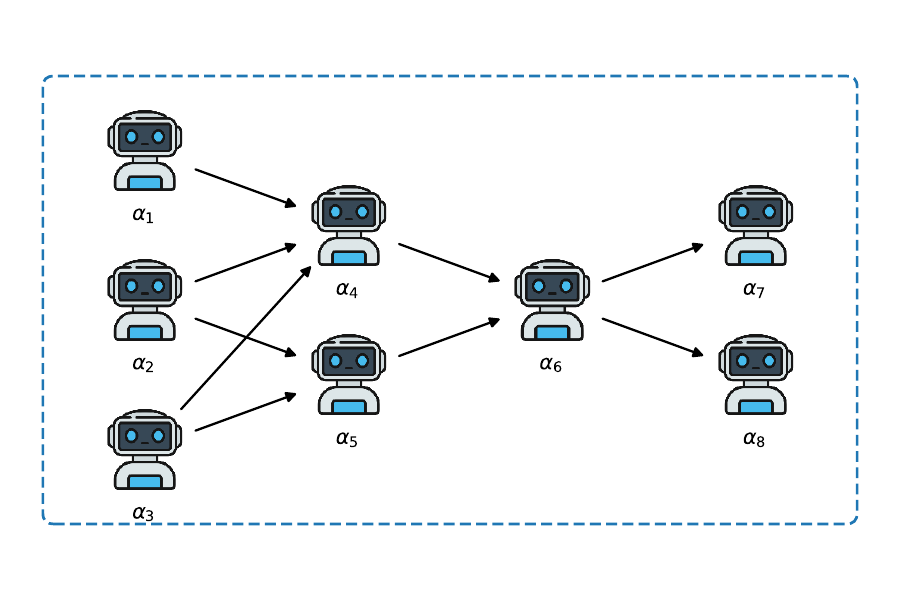}
    \caption{Workflow Size, $N=8$.}
    \label{fig:workflow-size}
\end{subfigure}
\begin{subfigure}[b]{0.33\textwidth}
    \includegraphics[width=\textwidth]{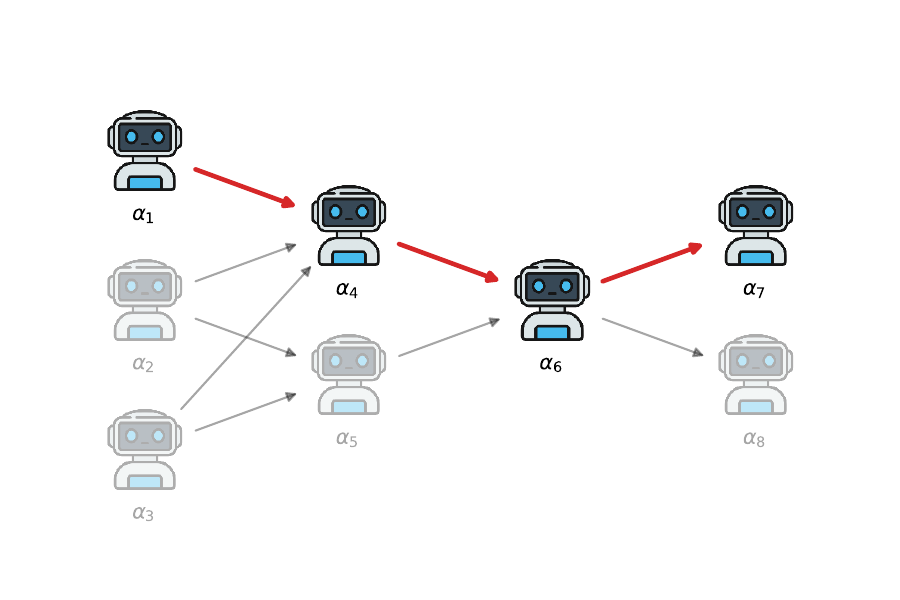}
    \caption{Workflow Depth, $d(G)=3$.}
    \label{fig:workflow-depth}
\end{subfigure}
\begin{subfigure}[b]{0.33\textwidth}
    \includegraphics[width=\textwidth]{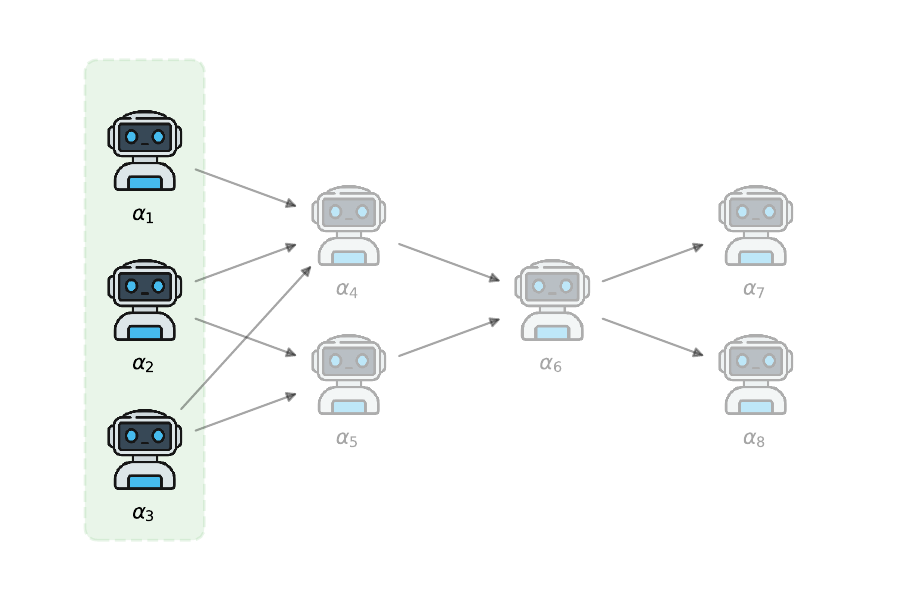}
    \caption{Workflow Width, $w(G)=3$.}
    \label{fig:workflow-width}
\end{subfigure}
\caption{An illustrative workflow with $N=8$, $d(G)=3$, $w(G)=3$ with each figure highlighting one structural property: (a) size, (b) depth, the longest entry-to-exit path and (c) width, the widest depth level.}
\label{fig:workflow_structure}
\end{figure*}

\section{Experimental Setup}
\label{sec:experimental-setup}

This section describes the experimental setup and methodology to compare the four schedulers in Section~\ref{sec:framework}. We present (i) the synthetic benchmark workflow generation for evaluation; (ii) the performance measures to inspect scheduling quality from the workflow scheduling, multi-objective and system perspectives; and (iii) the Wilcoxon rank-sum statistical significance test.

\subsection{Benchmark Workflows}

\begin{figure}[htbp]
\centerline{\includegraphics[width=0.50\textwidth]{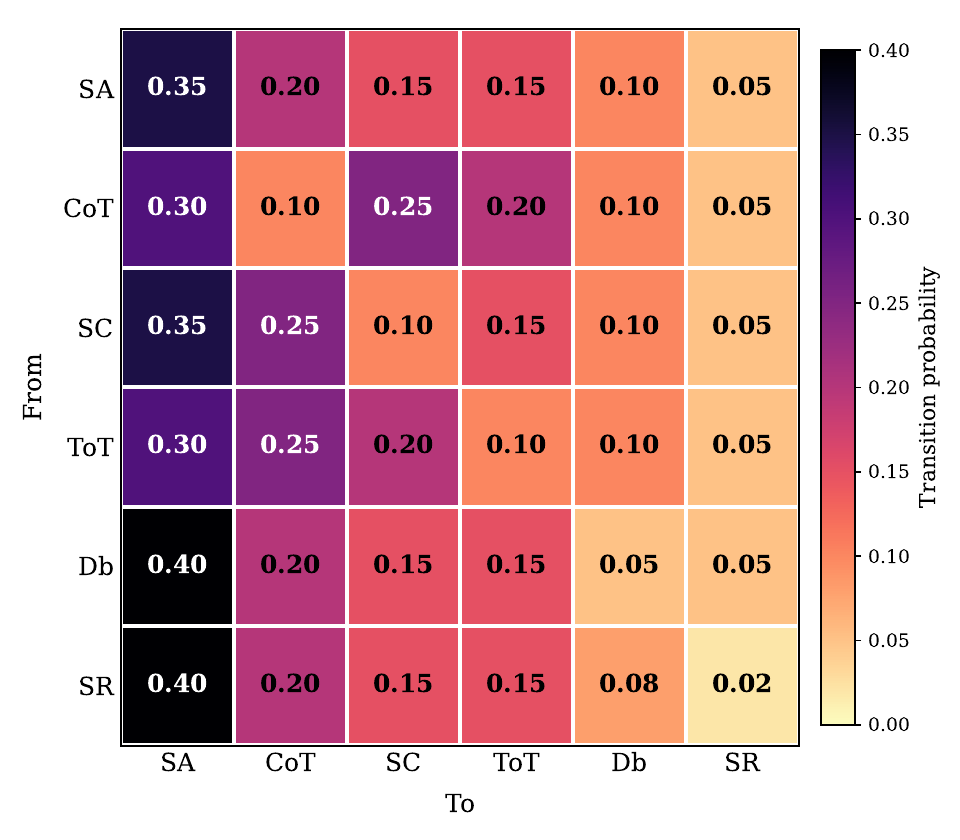}}
\caption{Transition-probability heatmap of the first-order Markov chain where rows are the current pattern and columns are the next. SA is the most likely next pattern regardless of the current.}
\label{fig:transition}
\end{figure}

Every workflow used in our experiments is generated by the parametric and synthetic workflow generator \cite{mawge} so that the workflow structure can be varied systematically and independently. Given a target size $N$, depth $d(G)$ and width $w(G)$ satisfying the feasibility inequality in Section~\ref{sec:workflow-model}, the generator constructs a layered DAG with exactly $d(G) + 1$ depth levels with the widest level of $w(G)$ agents and wires every agent in an interior level to at least one predecessor in the level above and at least one successor in the level below (see Figure~\ref{fig:workflow_structure}). In-degree and out-degree of each agent are drawn uniformly from $[1, 5]$. Each agent is additionally injected with up to five tool invocations. Every tool $\theta \in \Theta(\alpha)$ an agent invokes can be an internal tool with per-invocation latency $\ell_\theta \in [0.05, 2]$ seconds or an external tool with $\ell_\theta \in [0.5, 10]$ seconds, each invoked $n_\theta \in [1, 5]$ times. This topology construction lets three parameters $N$, $d(G)$ and $w(G)$ be set directly and independently. The generator also supports composing a workflow directly from the six canonical reasoning patterns of single-agent, chain-of-thought, self-consistency, tree-of-thought, debate and self-refine, ordered by the pattern-transition Markov chain \cite{mawge} to generate more semantically realistic agentic workflows (see Figure~\ref{fig:transition} for exact probabilities). Unless otherwise stated, experiments use the workflow size $N = 50$, depth $d(G) = 15$ and width $w(G) = 8$. Refer to Figure~\ref{lst:workflow_json_full} in Appendix~\ref{sec:workflow_json} for an illustrative workflow JSON.

\subsection{Performance Metrics}

We evaluate the schedulers with two levels of granularity as the solution-level metrics for the quality of a single scheduling solution and the population-level metrics for the quality of an entire population of solutions. These metrics are outlined as follows:

\subsubsection{Objectives} All scheduling solutions are primarily evaluated and analyzed on three optimization objectives of makespan, cost and utilization, as described in Section~\ref{sec:problem}. 

\subsubsection{Resource Load Imbalance} For a solution $\sigma$, let $R(\sigma)$ be the set of cloud resources it uses and let $T_\mathrm{busy}(\sigma, r)$ denote the total busy time of resource $r \in R(\sigma)$ under $\sigma$. Then, the resource load imbalance can be defined as:
\[
\frac{\displaystyle\max_{r \in R(\sigma)} T_\mathrm{busy}(\sigma, r) \;-\; \min_{r \in R(\sigma)} T_\mathrm{busy}(\sigma, r)}{\dfrac{1}{|R(\sigma)|}\displaystyle\sum_{r \in R(\sigma)} T_\mathrm{busy}(\sigma, r)}
\]
where zero imbalance indicates every resource carries an equal share of the workload. In this regard, larger imbalances imply that the total load is concentrated on a small subset of resources while other resources remain comparatively idle.

\subsubsection{Distinct Resource Count} Using $R(\sigma)$ as the set of cloud resources $\sigma$ uses, this metric counts the number of distinct cloud resources in $R(\sigma)$ as:
\[
|R(\sigma)|
\]
where it reveals the cloud footprint of a solution since every additional resource brings its own provisioning and deprovisioning overheads. Spreading the existing workload evenly on very few resources might improve cost and utilization.

\subsubsection{Number of Non-Dominated Solutions (NS)} This metric simply counts the number of non-dominated solutions in the front of a given population as $|F_0(\mathcal{P})|$. Larger values mean the scheduler finds a wider set of optimal solutions. However, interpretation of that metric might be occasionally misleading, especially when a better solution comes and dominates the existing ones in the front. 

\subsubsection{Schott's Spacing (SS)} Schott's spacing quantifies how evenly a front is spread \cite{schott1995}. Let $\hat{f}(\sigma) \in [0,1]^3$ be the min-max normalization of three objectives. For a front $F_0$, the nearest-neighbor distance of a solution $\sigma$ is:
\[
d_i \;=\; \min_{j \neq i} \; \bigl\| \hat{f}(\sigma_i) - \hat{f}(\sigma_j) \bigr\|_1
\]
Then, spacing is the sample standard deviation of these distances:
\[
\mathrm{SS}(F_0) \;=\; \sqrt{ \frac{1}{|F_0|-1} \sum_{i=1}^{|F_0|} \bigl( d_i - \bar{d} \bigr)^2 }, \qquad \bar{d} = \frac{1}{|F_0|}\sum_{i=1}^{|F_0|} d_i
\]
where lower values indicate a more uniformly spread front. $\mathrm{SS}(F_0) = 0$ means every solution in the front stays exactly $\bar{d}$ away from its nearest neighbor.

\subsubsection{Hypervolume (HV)} The hypervolume \cite{zitzler1998} measures the volume of objective space a front dominates. Using the same normalization factors as $\hat{f}(\cdot) \in [0,1]^3$ as spacing, let a reference point $\mathbf{r} = (1+\varepsilon,\, 1+\varepsilon,\, 1+\varepsilon)$ with a margin $\varepsilon$ on every axis so that even the worst solution still contributes positive volume. Hypervolume can be formally defined in the following way:
\[
\mathrm{HV}(F_0) \;=\; \Lambda \!\left( \bigcup_{\sigma \in F_0} \bigl[\hat{f}_M(\sigma), r_1\bigr] \times \bigl[\hat{f}_C(\sigma), r_2\bigr] \times \bigl[\hat{f}_U(\sigma), r_3\bigr] \right)
\]
where $\Lambda(\cdot)$ denotes three-dimensional Lebesgue measure (i.e., volume); and $\hat{f}_M(\sigma), \hat{f}_C(\sigma), \hat{f}_U(\sigma) \in [0,1]$ are the normalized objectives. For this metric, larger values indicate a front that dominates more objective space and lies closer to the ideal point.

\subsubsection{Pareto-Optimal Fronts} Solutions are projected onto three two-dimensional objective spaces as makespan--cost, makespan--utilization and cost--utilization; and one three-dimensional objective space as makespan--cost--utilization. These projections enable the achievable optimization surface to be visually inspected. 

\subsubsection{Scheduling Runtime} The wall-clock computational cost of running a scheduler is the elapsed time between the start and end times. It counts from the invocation of the scheduler to the return of the final solutions. Note that the scheduling runtime is distinct from the makespan of the scheduling solution the scheduler produces. 

\begin{table}[htbp]
\centering
\caption{Default experimental parameters and their corresponding values used across all the experiments unless otherwise stated.}
\label{table:parameters}
\begin{tabular}{ll}
\toprule
\textbf{Parameter} & \textbf{Value} \\
\midrule
Workflow size, $N$ & 50 \\
Workflow depth, $d(G)$ & 15 \\
Workflow width, $w(G)$ & 8 \\
\midrule
Cloud resource pool size, $|\mathcal{R}|$ & 60 \\
Resource-pool composition & CPU + GPU \\
vCPU weight, $w_v$ & 1.0 \\
vGPU weight, $w_g$ & 8.0 \\
Provisioning delay (mean / std) & 100 s / 10 s \\
Deprovisioning delay (mean / std) & 10 s / 2 s \\
\midrule
Population size, $\lambda$ & 50 \\
Generation count & 5000 \\
Mutation rate & 0.1 \\
Tournament size & 10 \\
\midrule
Independent runs, $n$ & 20 \\
\bottomrule
\end{tabular}
\end{table}

\begin{table*}[htbp]
\centering
\caption{Overall scheduling performance of the four schedulers, averaged over 20 independent runs with the improvement over RAND and statistical significance of * $p<0.05$.}
\label{table:overall_performance}
\begin{tabular}{lrrrr}
\toprule
\textbf{Metric} & \textbf{RAND} & \textbf{MO-HEFT} & \textbf{W-GA} & \textbf{NSGA-II} \\
\midrule
Makespan (s) $\downarrow$ & 2911.37 & 2706.69 (+7.0\%*) & 2731.46 (+6.2\%*) & 2663.27 (+8.5\%*) \\
Cost (\$) $\downarrow$ & 113.60 & 125.49 (-10.5\%) & 46.46 (+59.1\%*) & 6.17 (+94.6\%*) \\
Utilization $\uparrow$ & 30.39\% & 49.84\% (+64.0\%*) & 35.97\% (+18.4\%*) & 50.69\% (+66.8\%*) \\
Resource Imbalance $\downarrow$ & 2.52 & 1.49 (+40.6\%*) & 1.97 (+21.7\%*) & 1.63 (+35.4\%*) \\
Non-dominated Solutions $\uparrow$ & 12.30 & 21.95 (+78.5\%*) & 8.50 (-30.9\%*) & 25.20 (+104.9\%*) \\
Spacing (Schott's) $\downarrow$ & 0.06 & 0.07 (-12.0\%) & 0.03 (+47.7\%*) & 0.10 (-59.9\%*) \\
Hypervolume $\uparrow$ & 0.30 & 0.77 (+154.8\%*) & 0.56 (+85.7\%*) & 0.86 (+185.1\%*) \\
\midrule
Computation Time (s) & 0.0007 & 203.5308 & 73.5614 & 96.2599 \\
\bottomrule
\end{tabular}
\end{table*}

\subsection{Statistical Analysis}

All experiments are conducted on a single machine of MacBook Air with Apple M2 chip, 8 GB memory and 8 CPU cores. We run every scheduler independently $20$ times with freshly generated workflow instance and random seed for every run. Once a workflow is generated, all schedulers are evaluated on this workflow. We collect the performance metrics per run and report their results with their mean and standard deviations across these 20 runs. With independent measures, our goal is to eliminate the impact of randomness available in evolutionary computation and catch the typical behaviors of schedulers across distinct workflow instances, rather than a single specific draw. The RAND scheduler is just used as the baseline algorithm. The percentage improvements of the other schedulers with respect to RAND are reported where the positive percentage means better than the baseline and vice versa. We perform the statistical significance test via the Wilcoxon rank-sum test and report significance at the conventional $p < 0.05$ threshold. Refer to Table~\ref{table:parameters} for the list of the important parameters with their corresponding values used throughout the experimental setup. 

\section{Experimental Results}
\label{sec:experimental-results}

This section reports the experimental results and the key findings. We discuss (i) the overall scheduling performance of the schedulers; (ii) their resulting Pareto-optimal fronts; (iii) the effects of workflow topology; and (iv) effects of heterogeneous cloud resources.

\subsection{Overall Scheduling Performance}

Table~\ref{table:overall_performance} reports the makespan, cost, utilization and Pareto-optimal front quality of all four schedulers at the default configuration of Table~\ref{table:parameters} as the average of 20 independent runs and Wilcoxon rank-sum significance. The first and most critical observation is that \textbf{NSGA-II is the strongest scheduler that outperforms the other schedulers on most metrics by improving every objective.} It lowers makespan by $8.5\%$ to $2663.27$ seconds, lowers cost by $94.6\%$ from \$113.60 to \$6.17 and raises utilization by $66.8\%$ from $30.39\%$ to $50.69\%$ while also improving resource imbalance by $35.4\%$, the number of non-dominated solutions by $104.9\%$ and hypervolume by $185.1\%$. No other scheduler except W-GA improves makespan, cost and utilization at the same time.

\textbf{Cost is the objective that the schedulers separate most and makespan the least.} Cost spans over twenty-fold across the four schedulers from NSGA-II's \$6.17 to MO-HEFT's \$125.49. On the other hand, the schedulers except RAND stay within about $3\%$ of each other on makespan from $2663.27$ to $2731.46$ seconds. Figure~\ref{fig:makespan_scalability} shows that the inference part of each agent scaling with the resource it runs on is narrower than the tool latency, which leaves little room to minimize makespan. The hourly price of a resource, by contrast, varies widely across Table~\ref{tab:cloud_resources} and Table~\ref{tab:gpu_resources}. Hence, the search moves cost and utilization the most. \textbf{MO-HEFT reaches a high utilization, but without reaching it economically.} Its utilization of $49.84\%$ is close to NSGA-II's $50.69\%$ and it has the lowest resource imbalance at $1.49$. However, its cost of \$125.49 is the worst of all four schedulers. \textbf{W-GA improves cost and utilization but narrows the front.} It reduces cost by $59.1\%$ and raises utilization by $18.4\%$, but its count of non-dominated solutions falls to $8.50$ with a Schott's spacing of $0.03$.

\subsubsection{Generation Count} Figure~\ref{fig:convergence} in Appendix~\ref{sec:appendix_convergence} traces the best makespan, cost and utilization found by W-GA and NSGA-II over 5000 generations of a single run. \textbf{Cost improves earliest and most sharply for both schedulers.} W-GA reduces the best cost from about \$123.00 to about \$33.00 within the first thirty generations and NSGA-II reduces it from about \$146.00 to about \$25.00 within the first two hundred. The initial population assigns every agent to an independent uniformly-random resource and therefore contains many expensive instances that need to be removed. After that, both schedulers continue to lower cost slowly where W-GA reaches \$23.70 and NSGA-II reaches \$4.88 at generation 5000. \textbf{Makespan converges early and improves little afterwards.} W-GA settles around 2575 seconds by generation 600 and improves only to 2570 seconds for the remaining 4400 generations. NSGA-II settles around 2735 seconds by generation 1100 and later brings it to 2707 seconds. 

\textbf{Utilization is the slowest objective to converge for NSGA-II.} W-GA raises utilization from about 34\% to about 42\% by generation 1350 and then remains flat. NSGA-II instead climbs continuously, passing 50\% near generation 750, reaching 55\% near generation 2350 and completing at 56.5\%. \textbf{NSGA-II keeps discovering improvements later than W-GA.} The utilization of W-GA stops changing after roughly generation 1350 while its cost and makespan improve intermittently where the last of which occurs near generation 3750. By contrast, NSGA-II records a makespan step near generation 4200 and utilization gains past generation 4600. Selection of generation cut point forms a trade-off between scheduling quality and runtime.

\subsubsection{Temporal Scheduling Overhead} Figure~\ref{fig:scheduler_runtime} in Appendix~\ref{sec:appendix_computation_time} reports the wall-clock scheduling time of the four schedulers in MO-SAIC while Figure~\ref{fig:temporal_overhead} specifically focuses on NSGA-II for the growing population and workflow size. \textbf{RAND is trivially fast with 0.0007 seconds in total.} This is well expected since RAND performs only a single dependency-aware topological sort followed by a random resource draw. On the other hand, MO-HEFT, W-GA and NSGA-II take $203.53$, $73.56$ and $96.26$ seconds, respectively. \textbf{MO-HEFT is the most computationally expensive scheduler.} For each of the $N=50$ agents, it extends every $\lambda=50$ solution with every $|\mathcal{R}|=60$ default resource choice, producing $\lambda|\mathcal{R}|=3000$ candidate extensions that must be reduced back down to $\lambda$ solutions through the non-dominated sorting and crowding-distance operators.

\textbf{The heaviest computational overhead of MO-HEFT does not mean consistently superior scheduling quality.} According to Table~\ref{table:overall_performance}, MO-HEFT delivers a $7.0\%$ makespan gain over RAND and a worse cost of $\$125.49$ in return for significant gains in utilization. In contrast, NSGA-II achieves significant improvements over RAND, making it favorable for the balance between computational overhead and scheduling quality. \textbf{Applying non-dominated sorting of NSGA-II has a measurable overhead over weighted-sum of W-GA.} This is especially visible between W-GA's $73.5614$ seconds and NSGA-II's $96.2599$ seconds while they both run the identical population size and generation count. While W-GA employs a single weighted-sum fitness for solutions, NSGA-II performs non-dominated sorting and crowding distance at every generation. 

\subsection{Pareto Frontier Analysis}

This section presents three two-dimensional space projections of makespan--cost, makespan--utilization and cost--utilization from the three-dimensional objective space in Figure~\ref{fig:pareto_frontiers_all} for visual inspection.

\subsubsection{Makespan--Cost Frontier} \textbf{The scheduling solutions in Table~\ref{table:overall_performance} are distributed more widely along the cost axis than along the makespan axis.} Cost ranges over twenty-fold across the four schedulers, from NSGA-II's \$6.17 to MO-HEFT's \$125.49. However, makespan varies within about $3\%$ across the schedulers except RAND, from NSGA-II's $2663.27$ to W-GA's $2731.46$ seconds, with RAND higher at $2911.37$ seconds. Notably, NSGA-II dominates the other schedulers by finding the lowest cost and the lowest makespan. A spacing pattern is also visible in Figure~\ref{fig:pareto-nsga2} where solutions are clustered toward the low-cost and low-makespan corner, unlike those in Figure~\ref{fig:pareto-random} in Appendix~\ref{sec:appendix_pareto_frontiers}.

\subsubsection{Cost--Utilization Frontier} \textbf{Two most utilization-optimal schedulers act differently on the cost axis.} MO-HEFT and NSGA-II reach roughly the same utilization levels of $49.84\%$ and $50.69\%$, respectively. However, MO-HEFT's cost of \$125.49 is twenty times higher than NSGA-II's \$6.17, showing that MO-HEFT does not use resources economically. It suggests that reaching a utilization level offers no guarantee of reaching it economically, as seen in Figure~\ref{fig:pareto-moheft} and Figure~\ref{fig:pareto-nsga2} in Appendix~\ref{sec:appendix_pareto_frontiers}. Moreover, W-GA improves on RAND by raising utilization from $30.39\%$ to $35.97\%$ while reducing cost from \$113.60 to \$46.46.

\subsubsection{Makespan--Utilization Frontier} \textbf{A richer front does not necessarily mean a more evenly-distributed trade-off for makespan--utilization.} NSGA-II carries the most non-dominated solutions of the four schedulers with $25.20$ against RAND's $12.30$. However, its Schott's spacing of $0.10$ is worse than RAND's $0.06$, indicating that its larger front is not uniformly distributed, as seen in Figure~\ref{fig:pareto-nsga2}. W-GA shows the opposite pattern where it has fewer non-dominated solutions than even RAND, $8.50$ against $12.30$ while it has the tightest spacing of all four schedulers at $0.03$. Its weighted-sum fitness directs solutions toward a narrow and tightly-clustered region of the makespan--utilization space.

\subsection{Influence of Workflow Topology on NSGA-II}

This section evaluates the influence of workflow topology including size, depth and width on NSGA-II scheduling performance. 

\subsubsection{Workflow Size} Table~\ref{table:workflow_size} reports mean experimental results over 20 independent runs as the workflow size $N$ grows from 25 to 120 agents. \textbf{Cost grows faster than makespan as workflow size increases.} From $N=25$ to $N=120$, makespan rises by only about a third from $2292.91$ to $3057.72$ seconds while cost shows a dramatic rise of 44-fold from \$1.2469 to \$55.2051. Since workflow depth is fixed, the additional agents are included within the existing dependency levels rather than lengthening the critical path. Hence, the growing workload is executed largely in parallel, not affecting makespan to a great degree. In contrast, cost, as the sum of billing durations, accumulates roughly with each agent added and grows correspondingly faster than makespan. 

\textbf{NSGA-II absorbs the growing workload by recruiting more distinct resources at the expense of utilization and load imbalance.} The number of distinct resources grows more than threefold from $13.39$ at $N=25$ to $44.01$ at $N=120$ while utilization gradually falls from $56.93\%$ to $40.20\%$ and resource imbalance worsens from $1.3891$ to $2.1362$. A larger $N$ causes more agents at the same dependency levels to be ready at once and NSGA-II spreads them across a wider set of resources to avoid delays and keep makespan small. However, each resource comes with its hourly-rounded cycle that rarely aligns with how long that agent is actually busy, which lowers the average busy-time per resource as the total amount of work grows. Therefore, workflow size trades utilization and resource imbalance for bounded makespan growth.

\textbf{Increase in hypervolume does not necessarily reflect improvement with a stronger front.} Hypervolume climbs from $0.7466$ at $N=25$ to $0.9979$ at $N=100$, which would suggest an improvement on its own at the first glance. However, a larger workflow also widens the achievable range for makespan, cost and utilization, which can raise the hypervolume without an actual improvement. At the same time, the number of non-dominated solutions declines from $32.50$ at $N=25$ to roughly $21$ to $24$ for $N \geq 75$ and Schott's spacing worsens from $0.0843$ to $0.1093$. This suggests that NSGA-II yields a sparser and less evenly-distributed front. \textbf{Workflow size also inflates the computational cost of scheduling itself.} Figure~\ref{fig:temporal_overhead} in Appendix~\ref{sec:appendix_computation_time} reports the wall-clock computation time with the increasing workflow size. As shown, the runtime grows with $N$ since every additional agent lengthens the solution length and adds to the simulation complexity for every candidate solution at every generation. When evaluated together with the degrading multi-objective performance, workflow size is the most influential topology parameter by placing the heaviest demand on both scheduling quality and scheduling runtime scalability, among the three topology parameters considered.

\subsubsection{Workflow Depth}

Table~\ref{table:workflow_depth} reports how NSGA-II scheduling performance changes as depth $d(G)$ grows from 15 to 42 while workflow size $N=50$ and width $w(G)=8$ stay fixed. Note that $d(G)=42$ is the boundary value $w(G)+d(G)=N$ where it forces almost the entire workflow into a single chain with only one level with the full width $w(G)=8$. \textbf{Makespan grows while cost declines as workflow depth increases.} Makespan more than doubles from $2663.27$ seconds at $d(G)=15$ to $6220.51$ seconds at $d(G)=42$ while cost falls from \$6.17 to \$4.82 in the same range. In contrast to workflow size, depth lengthens the critical path where every additional level is a sequential dependency that cannot be parallelized regardless of how many resources are available. Hence, makespan scales with the number of levels. Cost falls since a narrower workflow needs less concurrency with fixed amount of work at any one time and schedulers can reuse already-active resources across consecutive levels of agents, rather than opening new billing cycles for additional resources.

\textbf{There is a small decline in both distinct resource count and utilization.} Distinct resources slightly fall from $27.00$ at $d(G)=15$ to $25.77$ at $d(G)=42$ while utilization declines from $50.69\%$ to $45.94\%$. Resource imbalance rises correspondingly from $1.6284$ at $d(G)=15$ to a peak of $1.9855$ at $d(G)=35$ since the wide $w(G)=8$ level of agents is spread across several resources while the remaining narrow levels are serialized onto a small set of heavily-reused resources. Finally, it settles at $1.9307$ at $d(G)=42$ as the workflow approaches a single chain. \textbf{Hypervolume also falls as depth grows.} Hypervolume declines from $0.8634$ at $d(G)=15$ to a low of $0.7482$ at $d(G)=35$, contrary to the increasing hypervolume observed for growing workflow size. Here, workflow size $N$ stays constant regardless of depth, so it does not enlarge achievable range of makespan, cost and utilization. In addition, heavily-serialized dependencies leave less freedom for a solution to trade makespan against cost and utilization, which compresses non-dominated solutions into a smaller region of objective space. This compression is also visible in Schott's spacing that decreases from $0.0998$ to $0.0724$ between $d(G)=15$ and $d(G)=35$. 

\subsubsection{Workflow Parallelism}

Table~\ref{table:workflow_width} in Appendix~\ref{sec:appendix_workflow_ablations} reports how NSGA-II scheduling performance changes as width $w(G)$ grows. Width sets the largest number of agents that can share a single dependency level. A wider workflow packs the same 50 agents into fewer and wider levels of simultaneous agents. \textbf{Makespan decreases slowly at the widest settings.} It stays near $2650$ seconds for $w(G)$ up to 15 and falls to $2491.95$ and $2347.12$ seconds at $w(G)=25$ and $35$. Since the depth is fixed, the critical path spans fifteen dependency levels and the achievable makespan is limited. \textbf{Cost shows no trend with width.} It fluctuates between \$4.7648 and \$6.5181 with a variance that overlaps for all five settings.

\textbf{Utilization rises overall as the workflow becomes wider.} It grows from $48.15\%$ at $w(G)=5$ to $59.09\%$ at $w(G)=35$ while resource imbalance improves from $1.7168$ to $1.5288$ and the distinct resource count falls from $26.73$ to $24.19$. A larger width lets more agents be ready at the same moment and NSGA-II packs onto a compact set of resources that then accumulate more busy time per billing cycle. This differs from the effect of workflow size where extra agents are spread across more resources. However, here, the agent count does not change and consequently increasing parallelism is absorbed by denser packing. \textbf{The Pareto-front contracts at high width.} Hypervolume falls from $0.9025$ at $w(G)=5$ to $0.7647$ at $w(G)=25$ while the number of non-dominated solutions holds near $25$ to $28$ and Schott's spacing stays near $0.09$ to $0.10$.

\subsection{Influence of Cloud Resources on NSGA-II}

This section evaluates the influence of cloud resources including resource size and CPU/GPU composition on NSGA-II scheduling performance. 

\subsubsection{Resource Size} Table~\ref{table:cloud_size} in Appendix~\ref{sec:appendix_cloud_ablations} reports the way NSGA-II scheduling performance changes as the cloud resource pool grows from $|\mathcal{R}|=60$ to $180$ and $300$ while the workflow topology stays fixed. \textbf{Makespan is insensitive to the size of the resource pool.} It moves from $2663.27$ seconds at $|\mathcal{R}|=60$ to $2676.70$ at $|\mathcal{R}|=180$ and $2591.85$ at $|\mathcal{R}|=300$ without a visible trend. For the workflow size of $N=50$, available resources are sufficient and adding more does not reduce makespan. However, resource size might be a bottleneck for larger workflows. \textbf{A larger pool inflates cost and lowers utilization.} Cost rises from \$6.1716 at $|\mathcal{R}|=60$ to \$17.8053 at $|\mathcal{R}|=180$ and \$25.7325 at $|\mathcal{R}|=300$ while utilization falls from $50.69\%$ to $45.67\%$ and resource imbalance worsens from $1.6284$ to $1.8114$. The number of distinct resources also grows from $27.00$ to $39.88$. \textbf{The Pareto-front does not improve with a larger pool.} The number of non-dominated solutions rises only slightly from $25.20$ to $28.50$, Schott's spacing improves marginally from $0.0998$ to $0.0839$ and hypervolume stays near $0.86$ with no clear trend. Overall, enlarging the resource pool beyond the concurrency the workflow requires brings no benefit to makespan or to front quality while it measurably worsens cost and utilization.

\subsubsection{Environment Heterogeneity}

Table~\ref{table:resource_pool} in Appendix~\ref{sec:appendix_cloud_ablations} reports NSGA-II scheduling performance when the resource pool is restricted to CPU-only or GPU-only instances against the default CPU+GPU environment. \textbf{The GPU-only pool gives the lowest makespan.} Makespan falls to $2620.16$ seconds against $2774.33$ for CPU-only and $2663.27$ for CPU+GPU, about $6\%$ below the CPU-only setting. GPU acceleration raises $\mathrm{cp}(r)$ and shortens the inference term of every agent. \textbf{The GPU-only pool costs roughly ten times more.} Cost reaches \$64.7780 against \$6.7696 for CPU-only and \$6.1716 for CPU+GPU. GPU instances carry much higher hourly prices and an all-GPU pool leaves no cheap fallback. CPU+GPU reaches the lowest cost of the three because it can place most agents on cheap CPU instances and use GPUs only where they help, under our simplified computation power model. 

\textbf{CPU-only reaches the highest utilization and CPU+GPU has the most non-dominated solutions.} Utilization is $53.86\%$ for CPU-only, $50.69\%$ for CPU+GPU and $46.84\%$ for GPU-only since fast GPU execution leaves resources idle within their billing hour. The number of non-dominated solutions is $25.20$ for CPU+GPU against $23.00$ for CPU-only and $15.60$ for GPU-only and Schott's spacing is best for CPU+GPU at $0.0998$. Both resource classes together give NSGA-II the widest set of makespan--cost trade-offs to populate. The GPU-only front collapses once makespan is satisfied by hardware alone. Overall, the mixed CPU+GPU environment delivers near the GPU-only makespan at the CPU-only cost with the most non-dominated solutions.

\section{Further Discussions}
\label{sec:discussions}

MO-SAIC makes a number of design and implementation assumptions to keep the problem tractable. In this section, we discuss such further issues as well as potential future research directions:

\begin{enumerate}[leftmargin=13pt]
    \item \textbf{Synthetic workflows rather than realistic agentic traces.} The workflows in this paper are generated through our synthetic and parametric workflow generator \cite{mawge} by compiling several agentic reasoning patterns under a Markovian transition matrix. Admittedly, this generator enables us to vary workflow size, depth and width independently and study their impacts in isolation. However, it might not fully reflect how real-world organizations compose their workflows in practice. In the future, we aim to collect realistic agentic workflow traces and re-evaluate our schedulers against them. 

    \item \textbf{A simulated but not measured execution environment.} Agent and cloud characteristics including inference latency, tool latency, provisioning and deprovisioning delays and network bandwidth are modeled in this work, by using static AWS pricing and LLM catalogs and Gaussian noise, rather than real-world measurements taken from cloud providers. However, real environment execution might show variability from network congestion, cold-start delays and interference, which can affect makespan. Deploying and simulating the MO-SAIC schedulers on an actual cloud environment is a promising direction.

    \item \textbf{A simplified computation power unification for CPUs and GPUs.} Our computation power model combines vCPU and vGPU counts through a single fixed-weighted sum $\mathrm{cp}(r) = w_v v_r + w_g g_r$ where it treats every vGPU unit as equally powerful regardless of GPU generation. However, in reality, a T4 and an H200 differ for the same workload. A natural next step is to replace this linear abstraction with per-generation throughput multipliers for more realistic workflow simulations.

    \item \textbf{Abstracting away hardware compatibility.} MO-SAIC permits any agent to be assigned to any cloud resource, so a scheduler might place a large model onto a CPU-only instance whose memory could not actually hold the model weights. However, in practice, model weights impose minimum memory footprint that only certain resources can satisfy. Some models might require multiple GPUs together, rather than a single resource. Consequently, some of the solutions our schedulers generate as optimal might not be deployed on real infrastructure. Therefore, incorporating per-model memory requirements into the assignment constraints might be a critical future direction for a production-ready MO-SAIC.

    \item \textbf{A static and offline scheduling formulation.} This paper considers the multi-objective optimization problem as time-invariant over fixed workflows and fixed cloud resources. However, real deployments might face changes over time, such as software and hardware failures. In addition, agents might fail and retry in tool calling. The schedulers in this paper are not capable of reacting to these scenarios effectively. Therefore, extending our work to a dynamic multi-objective problem would be beneficial for more adaptive multi-objective scheduling. 

    \item \textbf{Set of optimization objectives.} MO-SAIC optimizes  makespan, cost and utilization objectives by remaining entirely agnostic to the correctness of model responses. However, scheduling depending on smaller models might degrade the quality of that agent contribution even if it reduces cost and latency. This trade-off is not addressed where introducing task success probability or response quality as a further objective might be taken into consideration. In addition, energy consumption might be also studied in the future together with the existing objectives.
\end{enumerate}

\section{Conclusion}
\label{sec:conclusion}

In this paper, we formulate LLM-based multi-agent workflow scheduling in cloud computing as a multi-objective optimization problem, jointly considering the minimization of makespan and cost and the maximization of resource utilization. For the given problem, we propose MO-SAIC, a multi-objective scheduling framework with a total of four schedulers. To evaluate the schedulers in MO-SAIC, we employ a parametric and synthetic multi-agent workflow generator to systematically vary workflow topology with respect to size, depth and width. We measure performance under the extensive test scenarios as: (i) their overall performance improvements relative to RAND; (ii) quality of their fronts; (iii) impacts of workflow topology; and (iv) impacts of cloud resources. The performance metrics we use span three main perspectives as: (i) objective functions, resource load imbalance, distinct resource count for the workflow scheduling perspective; (ii) number of non-dominated solutions, Schott’s spacing, hypervolume and visual Pareto-optimal fronts for the multi-objective perspective; and (iii) scheduling runtime for the system perspective. Lastly, we identify the current limitations of MO-SAIC and the potential future research directions. We expect that, rather than the model layer, the operational layer of artificial intelligence will be increasingly important in the future as the open-weight models close the performance gap with the proprietary ones, making the multi-objective multi-agent workflow scheduling a critical problem to address for large organizations. 

\clearpage
\onecolumn

\appendix

\section{LLM Model Performance Reference}
\label{sec:appendix_model_performance}

This appendix lists the performance characteristics of the agent models in terms of throughput and TTFT.

\begin{table*}[htbp]
\centering
\small
\caption{Large language models with reference performance characteristics of throughput and TTFT \cite{benchlmai}, assumed to be measured at a single-vCPU configuration. In this work, we deliberately treat the models as open-source and self-hosted on cloud resources so that they can be scheduled onto cloud resources.}
\label{tab:model_performance}
\begin{tabular}{@{}lrr@{\hspace{1cm}}lrr@{}}
\toprule
\textbf{Model} & \textbf{Tput} & \textbf{TTFT} & \textbf{Model} & \textbf{Tput} & \textbf{TTFT} \\
 & \textbf{(t/s)} & \textbf{(s)} & & \textbf{(t/s)} & \textbf{(s)} \\
\midrule
\multicolumn{3}{@{}l@{}}{\textsc{OpenAI Models}} & \multicolumn{3}{@{}l@{}}{\textsc{OpenAI Models}} \\
\addlinespace[0.05cm]
gpt-5.4-pro & 74 & 151.79 & o1 & 98 & 32.29 \\
gpt-5.4 & 74 & 151.79 & o3 & 118 & 5.38 \\
gpt-5.4-mini & 201 & 3.85 & o3-mini & 160 & 7.12 \\
gpt-5.4-nano & 191 & 3.64 & o3-pro & 27 & 84.93 \\
gpt-5.3-codex & 79 & 88.26 & o4-mini & 161 & 21.94 \\
gpt-5.2-codex & 123 & 87.34 & \multicolumn{3}{@{}l@{}}{\textsc{Anthropic Claude}} \\
\addlinespace[0.05cm]
gpt-5.2 & 73 & 130.34 & claude-4.1-opus & 29 & 1.66 \\
gpt-5.1 & 111 & 57.47 & claude-4.1-opus-thinking & 29 & 15.00 \\
gpt-5 & 83 & 36.28 & claude-4-sonnet & 40 & 1.33 \\
gpt-5-nano & 137 & 83.30 & claude-4.6-sonnet & 44 & 1.48 \\
gpt-5-mini & 86 & 65.32 & claude-4.5-opus & 46 & 1.01 \\
gpt-4.1 & 108 & 1.02 & claude-4.6-opus & 40 & 1.78 \\
gpt-4.1-nano & 181 & 0.63 & claude-3-haiku & 138 & 1.16 \\
gpt-4.1-mini & 80 & 0.76 & \multicolumn{3}{@{}l@{}}{\textsc{Google Gemini}} \\
\addlinespace[0.05cm]
gpt-4o & 131 & 0.81 & gemini-3.1-pro & 109 & 29.71 \\
gpt-4o-mini & 33 & 3.16 & gemini-3.1-flash-lite & 205 & 7.50 \\
gpt-4-turbo & 30 & 2.84 & gemini-3-pro & 109 & 32.65 \\
gpt-oss-120b & 262 & 0.79 & gemini-3-flash & 159 & 1.19 \\
gpt-oss-20b & 313 & 0.65 & & & \\
\addlinespace[0.1cm]
\midrule
\multicolumn{3}{@{}l@{}}{\textsc{Google Gemini}} & \multicolumn{3}{@{}l@{}}{\textsc{Mistral Models}} \\
\addlinespace[0.05cm]
gemini-2.5-pro & 117 & 21.19 & mistral-small-4 & 175 & 0.64 \\
gemini-2.5-flash & 221 & 0.50 & mistral-medium-3 & 57 & 1.20 \\
\multicolumn{3}{@{}l@{}}{\textsc{Meta Llama}} & mistral-large-3 & 48 & 1.04 \\
\addlinespace[0.05cm]
llama-4-scout & 128 & 0.70 & mistral-large-2 & 38 & 1.45 \\
llama-4-maverick & 121 & 0.95 & \multicolumn{3}{@{}l@{}}{\textsc{xAI Grok}} \\
\addlinespace[0.05cm]
llama-3.1-405b & 29 & 2.19 & grok-4.20 & 233 & 10.33 \\
\multicolumn{3}{@{}l@{}}{\textsc{Mistral Models}} & grok-4.3 & 209 & 12.36 \\
\addlinespace[0.05cm]
ministral-3-3b & 274 & 0.42 & grok-4.1-fast & 138 & 0.54 \\
ministral-3-8b & 182 & 0.52 & grok-4 & 54 & 15.60 \\
ministral-3-14b & 110 & 0.60 & grok-3-mini & 190 & 0.54 \\
\addlinespace[0.1cm]
\midrule
\multicolumn{3}{@{}l@{}}{\textsc{xAI Grok}} & \multicolumn{3}{@{}l@{}}{\textsc{Other Models}} \\
\addlinespace[0.05cm]
grok-code-fast-1 & 172 & 2.81 & glm-5 & 74 & 1.64 \\
\multicolumn{3}{@{}l@{}}{\textsc{Other Models}} & glm-4.7 & 82 & 1.10 \\
\addlinespace[0.05cm]
nemotron-3-super-100b & 367 & 0.71 & glm-4.7-flash & 95 & 0.91 \\
nemotron-3-nano-30b & 152 & 1.90 & glm-4.5 & 51 & 1.45 \\
deepseek-v3.2 & 35 & 3.75 & glm-4.5-air & 106 & 1.18 \\
deepseek-r1-distill-qwen-32b & 60 & 0.84 & lfm2-24b-a2b & 92 & 0.42 \\
mercury-2-inception & 789 & 3.88 & step-3.5-flash & 87 & 3.03 \\
ling-2.6-flash & 210 & 1.07 & minimax-m2.7 & 45 & 2.53 \\
nova-pro & 141 & 0.81 & minimax-m2.5 & 46 & 2.12 \\
mimo-v2-flash & 129 & 2.14 & kimi-k2.5 & 45 & 2.38 \\
qwen-3.5-397b & 96 & 2.44 & kimi-k2 & 43 & 1.51 \\
& & & phi-4 & 35 & 2.02 \\
& & & gemma-3-27b & 31 & 2.04 \\
\bottomrule
\end{tabular}
\begin{minipage}{\textwidth}
\end{minipage}
\end{table*}

\clearpage

\section{Cloud Resource Specifications}
\label{sec:appendix_cloud_resources}

This appendix presents the specifications of CPU-only and GPU-accelerated cloud resources on which agents are deployed and executed.

\begin{table*}[htbp]
\centering
\footnotesize
\caption{AWS EC2 CPU-Only instance specifications including virtual CPU count, network bandwidth and hourly on-demand pricing for five resource families (M1, M3, M4, M5, M5D) \cite{vantage}.}
\label{tab:cloud_resources}
\begin{tabular}{@{}lrrr@{\hspace{1cm}}lrrr@{}}
\toprule
\textbf{Instance Type} & \textbf{vCPUs} & \textbf{Network} & \textbf{Price} & \textbf{Instance Type} & \textbf{vCPUs} & \textbf{Network} & \textbf{Price} \\
 & & \textbf{(Gbps)} & \textbf{(\$/hour)} & & & \textbf{(Gbps)} & \textbf{(\$/hour)} \\
\midrule
\multicolumn{4}{@{}l@{}}{\textsc{M1 Family}} & \multicolumn{4}{@{}l@{}}{\textsc{M4 Family}} \\
\addlinespace[0.05cm]
m1.small & 1 & 10 & 0.044 & m4.large & 2 & 10 & 0.1 \\
m1.medium & 1 & 10 & 0.087 & m4.xlarge & 4 & 10 & 0.2 \\
m1.large & 2 & 10 & 0.175 & m4.2xlarge & 8 & 10 & 0.4 \\
m1.xlarge & 4 & 10 & 0.35 & m4.4xlarge & 16 & 10 & 0.8 \\
\multicolumn{4}{@{}l@{}}{\textsc{M3 Family}} & m4.10xlarge & 40 & 10 & 2.0 \\
\addlinespace[0.05cm]
m3.medium & 1 & 10 & 0.067 & m4.16xlarge & 64 & 25 & 3.2 \\
m3.large & 2 & 10 & 0.133 & \multicolumn{4}{@{}l@{}}{\textsc{M5 Family}} \\
\addlinespace[0.05cm]
m3.xlarge & 4 & 10 & 0.266 & m5.large & 2 & 10 & 0.096 \\
m3.2xlarge & 8 & 10 & 0.532 & m5.xlarge & 4 & 10 & 0.192 \\
\multicolumn{4}{@{}l@{}}{\textsc{M5 Family}} & m5.2xlarge & 8 & 10 & 0.384 \\
\addlinespace[0.05cm]
m5.4xlarge & 16 & 10 & 0.768 & m5.8xlarge & 32 & 10 & 1.536 \\
m5.12xlarge & 48 & 12 & 2.304 & m5.16xlarge & 64 & 20 & 3.072 \\
m5.24xlarge & 96 & 25 & 4.608 & & & & \\
\addlinespace[0.1cm]
\midrule
\multicolumn{4}{@{}l@{}}{\textsc{M5D Family}} & \multicolumn{4}{@{}l@{}}{\textsc{M5D Family}} \\
\addlinespace[0.05cm]
m5d.large & 2 & 10 & 0.113 & m5d.xlarge & 4 & 10 & 0.226 \\
m5d.2xlarge & 8 & 10 & 0.452 & m5d.4xlarge & 16 & 10 & 0.904 \\
m5d.8xlarge & 32 & 10 & 1.808 & m5d.12xlarge & 48 & 12 & 2.712 \\
m5d.16xlarge & 64 & 20 & 3.616 & m5d.24xlarge & 96 & 25 & 5.424 \\
\bottomrule
\end{tabular}
\vspace{0.1cm}
\begin{minipage}{\textwidth}
\end{minipage}
\end{table*}

\begin{table*}[htbp]
\footnotesize
\small
\caption{AWS EC2 GPU-accelerated instance specifications including virtual CPU count, virtual GPU count, network bandwidth and hourly on-demand pricing for ten NVIDIA GPU instance families from T4 through H200 (G4dn, G5, G6, G6e, G7, G7e, P4d, P4de, P5 and P5en) \cite{vantage}.}
\label{tab:gpu_resources}
\begin{tabular}{@{}lrrrr@{\hspace{1cm}}lrrrr@{}}
\toprule
\textbf{Instance Type} & \textbf{vCPUs} & \textbf{vGPUs} & \textbf{Network} & \textbf{Price} & \textbf{Instance Type} & \textbf{vCPUs} & \textbf{vGPUs} & \textbf{Network} & \textbf{Price} \\
 & & & \textbf{(Gbps)} & \textbf{(\$/hour)} & & & & \textbf{(Gbps)} & \textbf{(\$/hour)} \\
\midrule
\multicolumn{5}{@{}l@{}}{\textsc{NVIDIA T4 -- 16 GB/GPU}} & \multicolumn{5}{@{}l@{}}{\textsc{NVIDIA L40S -- 48 GB/GPU}} \\
g4dn.xlarge & 4 & 1 & 25 & 0.526 & g6e.xlarge & 4 & 1 & 20 & 1.861 \\
g4dn.2xlarge & 8 & 1 & 25 & 0.752 & g6e.2xlarge & 8 & 1 & 20 & 2.242 \\
\multicolumn{5}{@{}l@{}}{\textsc{NVIDIA A10G -- 24 GB/GPU}} & \multicolumn{5}{@{}l@{}}{\textsc{NVIDIA RTX PRO -- 32 GB/GPU}} \\
g5.xlarge & 4 & 1 & 10 & 1.006 & g7.2xlarge & 8 & 1 & 60 & 2.520 \\
g5.2xlarge & 8 & 1 & 10 & 1.212 & g7.4xlarge & 16 & 1 & 100 & 3.042 \\
g5.4xlarge & 16 & 1 & 25 & 1.624 & g7.8xlarge & 32 & 1 & 100 & 4.086 \\
g5.12xlarge & 48 & 4 & 40 & 5.672 & g7.24xlarge & 96 & 4 & 350 & 14.257 \\
g5.16xlarge & 64 & 1 & 25 & 4.096 & g7.48xlarge & 192 & 8 & 700 & 28.513 \\
g5.24xlarge & 96 & 4 & 50 & 8.144 & \multicolumn{5}{@{}l@{}}{\textsc{NVIDIA RTX PRO -- 96 GB/GPU}} \\
\multicolumn{5}{@{}l@{}}{\textsc{NVIDIA L4 -- 24 GB/GPU}} & g7e.2xlarge & 8 & 1 & 50 & 3.363 \\
g6.xlarge & 4 & 1 & 10 & 0.805 & g7e.4xlarge & 16 & 1 & 50 & 3.998 \\
g6.2xlarge & 8 & 1 & 10 & 0.978 & \multicolumn{5}{@{}l@{}}{\textsc{NVIDIA A100 -- 40 GB/GPU}} \\
g6.4xlarge & 16 & 1 & 25 & 1.323 & p4d.24xlarge & 96 & 8 & 400 & 21.958 \\
g6.8xlarge & 32 & 1 & 25 & 2.014 & \multicolumn{5}{@{}l@{}}{\textsc{NVIDIA A100 -- 80 GB/GPU}} \\
g6.12xlarge & 48 & 4 & 40 & 4.602 & p4de.24xlarge & 96 & 8 & 400 & 27.447 \\
g6.16xlarge & 64 & 1 & 25 & 3.397 & \multicolumn{5}{@{}l@{}}{\textsc{NVIDIA H100 -- 80 GB/GPU}} \\
g6.24xlarge & 96 & 4 & 50 & 6.675 & p5.4xlarge & 16 & 1 & 100 & 6.880 \\
g6.48xlarge & 192 & 8 & 100 & 13.350 & p5.48xlarge & 192 & 8 & 3200 & 55.040 \\
 & & & & & \multicolumn{5}{@{}l@{}}{\textsc{NVIDIA H200 -- 141 GB/GPU}} \\
 & & & & & p5en.48xlarge & 192 & 8 & 3200 & 63.296 \\
\bottomrule
\end{tabular}
\end{table*}

\clearpage

\section{A Workflow JSON}
\label{sec:workflow_json}

This appendix lists the per-agent attributes for the agentic workflow in Figure~\ref{fig:workflow_structure} in the format of JSON.

\begin{figure*}[htbp]
\centering
\begin{minipage}[htbp]{0.48\textwidth}
\begin{lstlisting}[style=json]
[
  {
    "id": "1", 
    "model": "gemini-2.5-pro", 
    "ttft": 21.19, 
    "tokens_per_sec": 117,
    "output_size": 175,
    "successors": ["4"],
    "tools": [{"is_internal": true, 
               "latency": 1.49, 
               "number_of_requests": 5}, ...]
  },
  {
    "id": "2", 
    "model": "gpt-5.4-pro", 
    "ttft": 151.79, 
    "tokens_per_sec": 74,
    "output_size": 438,
    "successors": ["4", "5"],
    "tools": [{"is_internal": true, 
               "latency": 0.59, 
               "number_of_requests": 2}, ...]
  },
  {
    "id": "3", 
    "model": "gpt-5-mini", 
    "ttft": 65.32, 
    "tokens_per_sec": 86,
    "output_size": 332,
    "successors": ["4", "5"],
    "tools": [{"is_internal": false, 
               "latency": 8.68, 
               "number_of_requests": 5}, ...]
  },
  {
    "id": "4", 
    "model": "mistral-large-3", 
    "ttft": 1.04, 
    "tokens_per_sec": 48,
    "output_size": 157,
    "successors": ["6"],
    "tools": [{"is_internal": false, 
               "latency": 6.65, 
               "number_of_requests": 5}, ...]
  },
\end{lstlisting}
\end{minipage}\hfill
\begin{minipage}[htbp]{0.48\textwidth}
\begin{lstlisting}[style=json]
  {
    "id": "5", 
    "model": "grok-3-mini", 
    "ttft": 0.54, 
    "tokens_per_sec": 190,
    "output_size": 187,
    "successors": ["6"],
    "tools": [{"is_internal": false, 
               "latency": 3.48, 
               "number_of_requests": 4}, ...]
  },
  {
    "id": "6", 
    "model": "gemma-3-27b", 
    "ttft": 2.04, 
    "tokens_per_sec": 31,
    "output_size": 254,
    "successors": ["7", "8"],
    "tools": [{"is_internal": false, 
               "latency": 1.36, 
               "number_of_requests": 1}, ...]
  },
  {
    "id": "7", 
    "model": "gpt-4o", 
    "ttft": 0.81, 
    "tokens_per_sec": 131,
    "output_size": 128,
    "successors": [],
    "tools": [{"is_internal": true, 
               "latency": 0.17, 
               "number_of_requests": 4}, ...]
  },
  {
    "id": "8", 
    "model": "gpt-4o", 
    "ttft": 0.81, 
    "tokens_per_sec": 131,
    "output_size": 200,
    "successors": [],
    "tools": [{"is_internal": true, 
               "latency": 1.45, 
               "number_of_requests": 3}, ...]
  }
]
\end{lstlisting}
\end{minipage}
\caption{Attributes of all eight agents for the workflow in Figure~\ref{fig:workflow_structure}. MO-SAIC schedulers consider these attributes while scheduling the agents onto the cloud resources. }
\label{lst:workflow_json_full}
\end{figure*}

\clearpage

\section{Wall-Clock Computation Time}
\label{sec:appendix_computation_time}

This appendix provides wall-clock scheduling overheads of schedulers in MO-SAIC.

\begin{figure*}[htbp]
\centerline{\includegraphics[width=.49\textwidth]{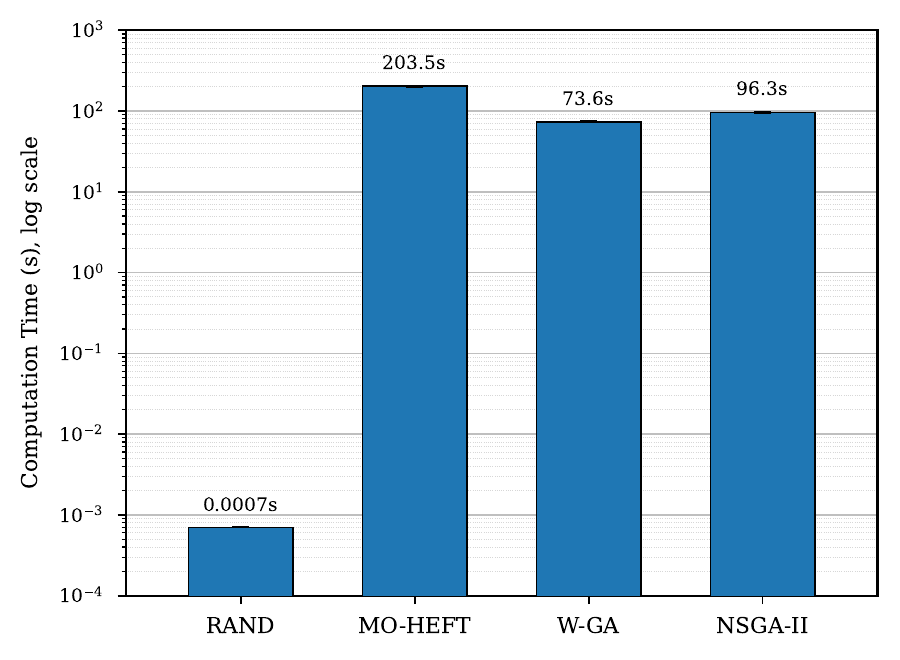}}
\caption{Computation time of the schedulers (mean $\pm$ std over 20 runs) at default configuration, shown on a log scale.}
\label{fig:scheduler_runtime}
\end{figure*}

\begin{figure*}[htbp]
\centerline{\includegraphics[width=0.80\textwidth]{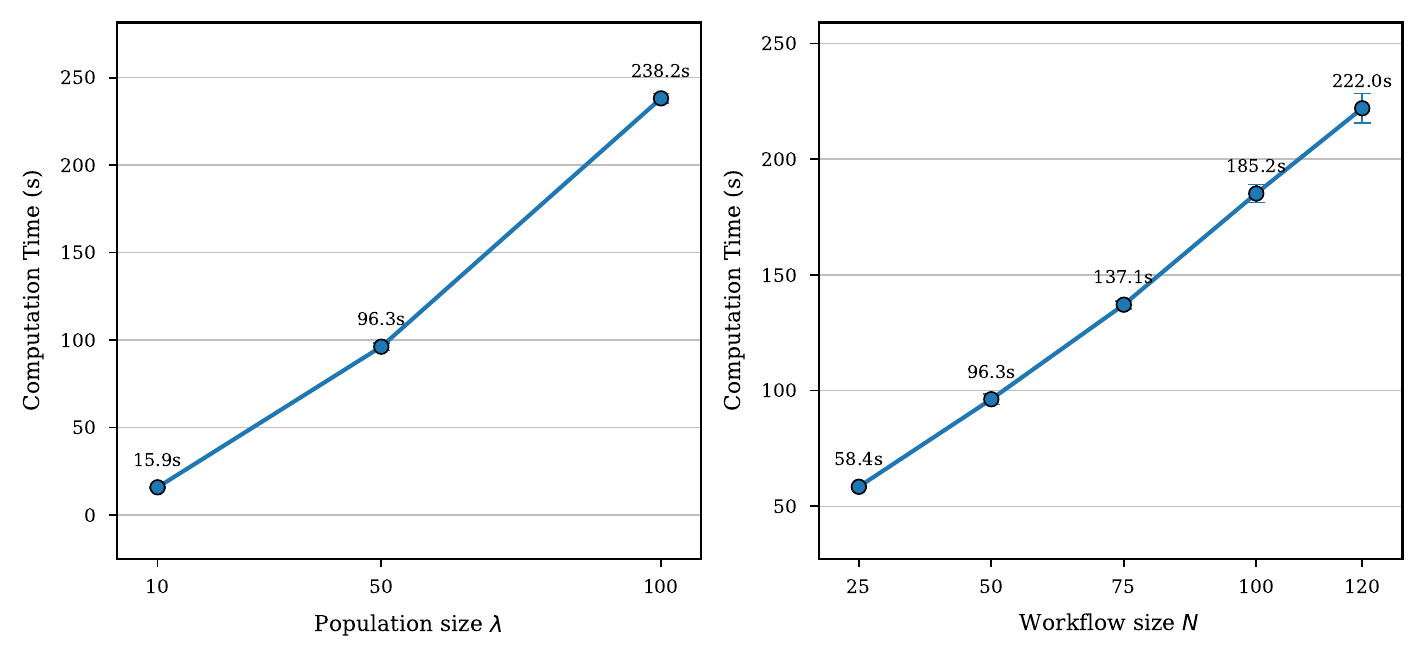}}
\caption{NSGA-II computation time (mean $\pm$ std over 20 independent workflows) as population size $\lambda$ (left) and workflow size $N$ (right) grow.}
\label{fig:temporal_overhead}
\end{figure*}

\clearpage

\section{Convergence of Genetic Algorithms}
\label{sec:appendix_convergence}

This appendix shows variations of the best makespan, cost and utilization values across all 5000 generations of a single run. This is only applicable to the generational schedulers including W-GA and NSGA-II.

\begin{figure*}[htbp]
    \centering
    \begin{subfigure}[b]{0.99\textwidth}
        \includegraphics[width=\textwidth]{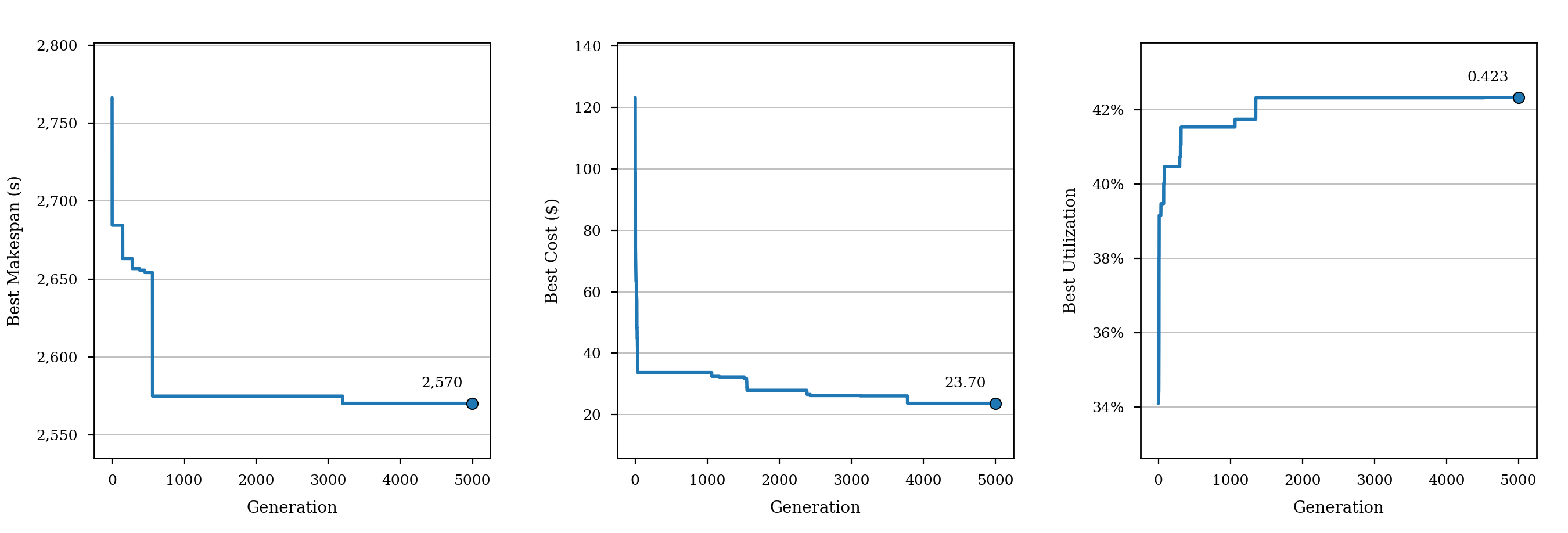}
        \caption{Convergence of W-GA over 5000 generations.}
        \label{fig:ga-convergence}
    \end{subfigure} \\
    \begin{subfigure}[b]{0.99\textwidth}
        \includegraphics[width=\textwidth]{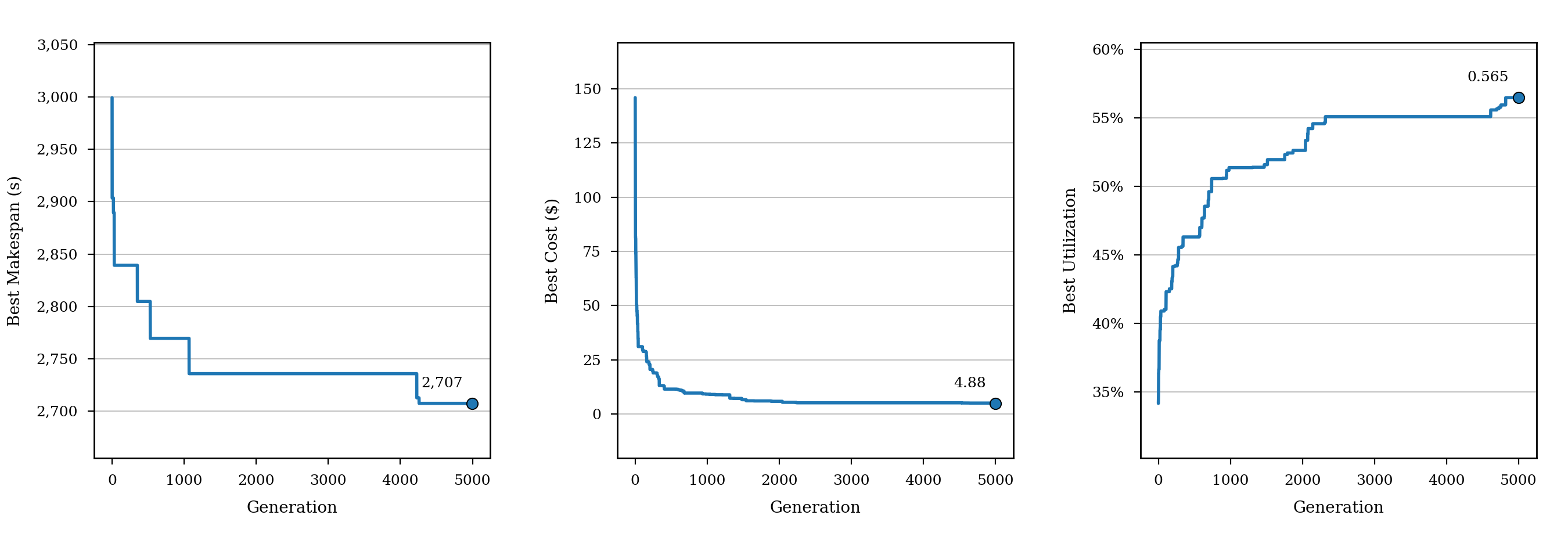}
        \caption{Convergence of NSGA-II over 5000 generations.}
        \label{fig:nsga2-convergence}
    \end{subfigure}
    \caption{Convergence of the W-GA and NSGA-II schedulers over 5000 generations on makespan, cost and utilization.}
    \label{fig:convergence}
\end{figure*}

\clearpage

\section{Two-Dimensional Pareto Frontiers}
\label{sec:appendix_pareto_frontiers}

This appendix shows the Pareto frontiers of the schedulers based on their final populations for a single run in two-dimensional projections.

\begin{figure*}[htbp]
    \centering
    \begin{subfigure}[b]{0.75\textwidth}
        \includegraphics[width=\textwidth]{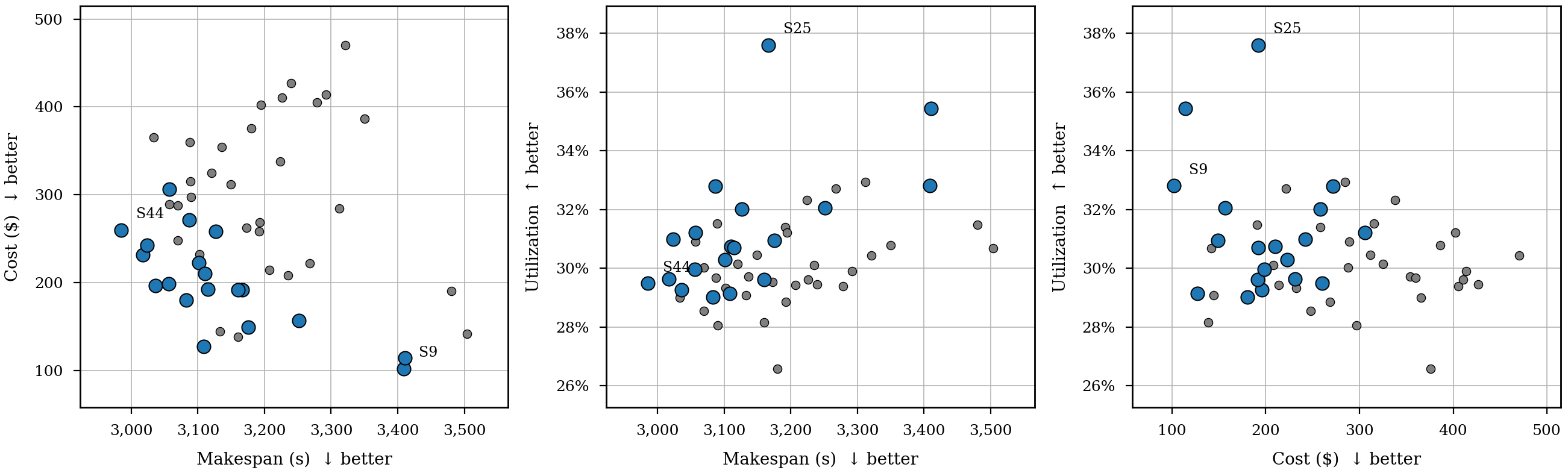}
        \caption{RAND: makespan--cost, makespan--utilization and cost--utilization projections.}
        \label{fig:pareto-random}
    \end{subfigure} \\
    \begin{subfigure}[b]{0.75\textwidth}
        \includegraphics[width=\textwidth]{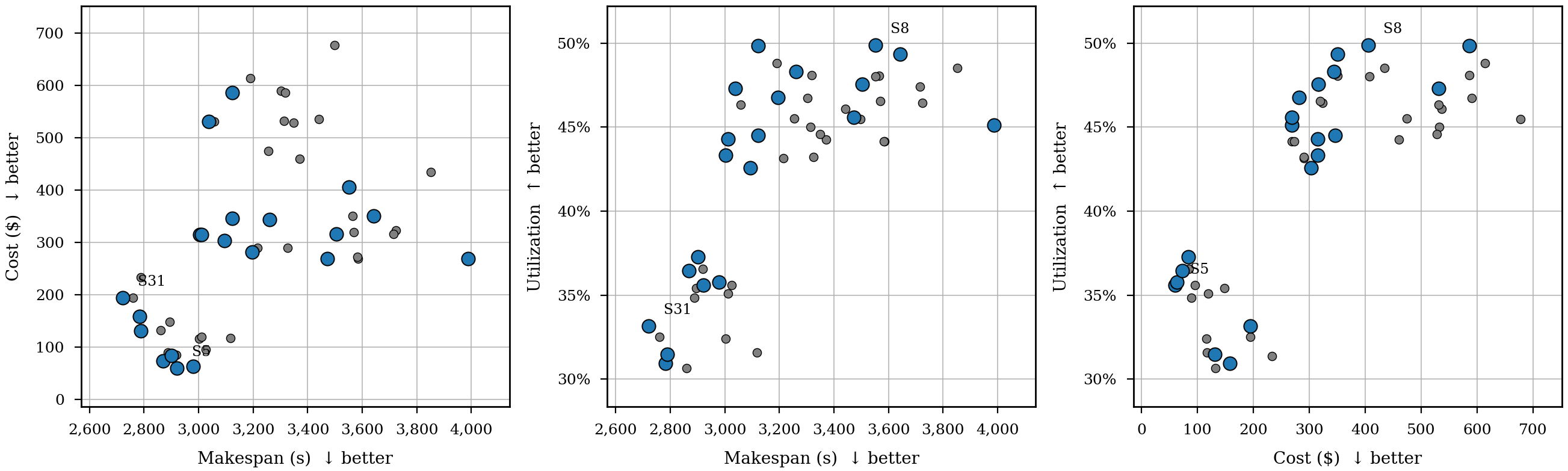}
        \caption{MO-HEFT: makespan--cost, makespan--utilization and cost--utilization projections.}
        \label{fig:pareto-moheft}
    \end{subfigure}
    \begin{subfigure}[b]{0.75\textwidth}
        \includegraphics[width=\textwidth]{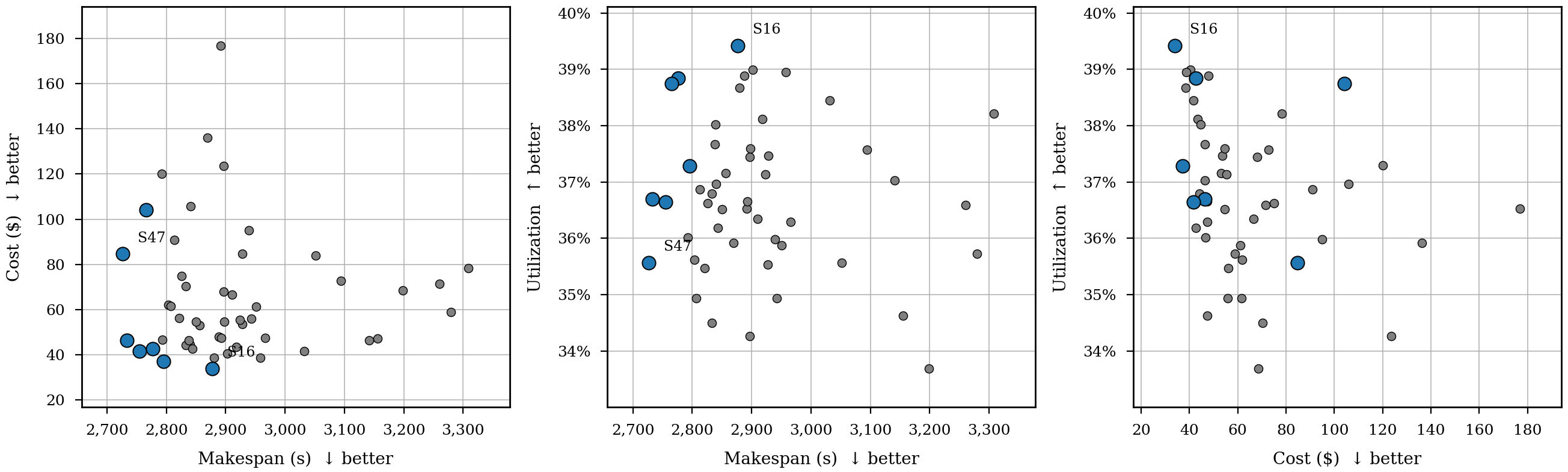}
        \caption{W-GA: makespan--cost, makespan--utilization and cost--utilization projections.}
        \label{fig:pareto-ga}
    \end{subfigure}
    \begin{subfigure}[b]{0.75\textwidth}
        \includegraphics[width=\textwidth]{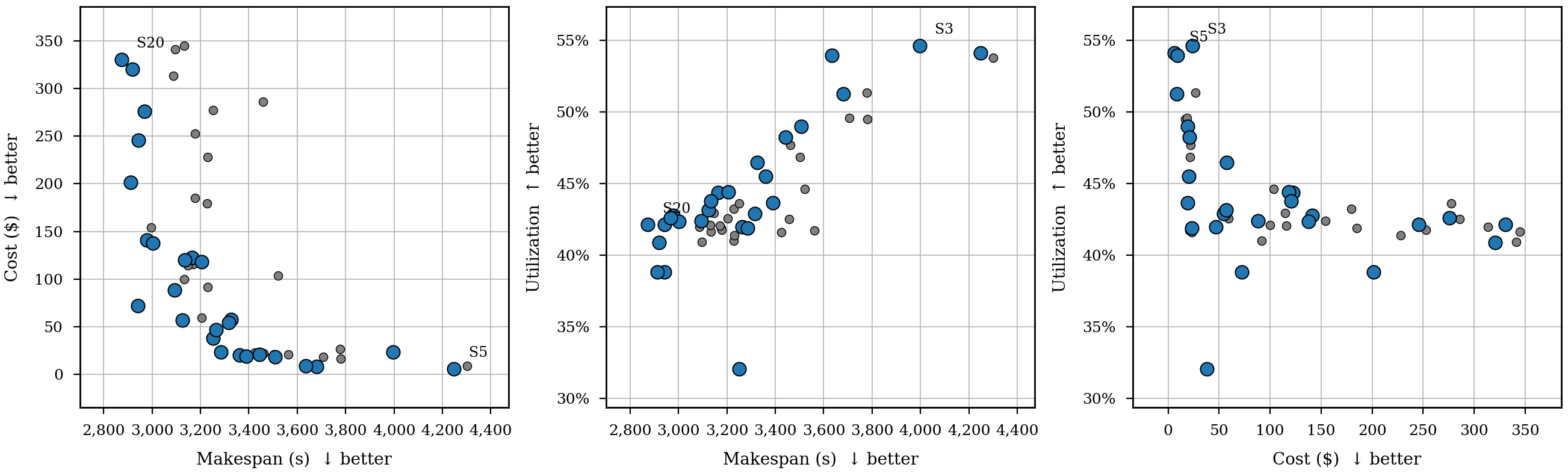}
        \caption{NSGA-II: makespan--cost, makespan--utilization and cost--utilization projections.}
        \label{fig:pareto-nsga2}
    \end{subfigure}
    \caption{Non-dominated (blue) and dominated (gray) solutions from the final populations of schedulers. Solutions are projected onto the three two-dimensional objective spaces.}
    \label{fig:pareto_frontiers_all}
\end{figure*}

\clearpage

\section{Workflow Topology}
\label{sec:appendix_workflow_ablations}

This appendix provides tables for the impacts of workflow topology such as size, depth and width on the NSGA-II performance.

\begin{table*}[htbp]
\centering
\caption{Impact of workflow size $N$ on NSGA-II scheduling quality (mean $\pm$ std over 20 independent workflows), $d(G)=15$, $w(G)=8$ and all other parameters at their defaults.}
\label{table:workflow_size}
\begin{tabular}{lrrrrr}
\toprule
\textbf{Workflow size $N$} & \textbf{25} & \textbf{50} & \textbf{75} & \textbf{100} & \textbf{120} \\
\midrule
Makespan (s) $\downarrow$ & $2292.91 \pm 132.04$ & $2663.27 \pm 148.18$ & $2871.44 \pm 180.93$ & $3035.66 \pm 120.95$ & $3057.72 \pm 152.35$ \\
Cost (\$) $\downarrow$ & $1.2469 \pm 0.1909$ & $6.1716 \pm 1.3416$ & $17.0122 \pm 2.4709$ & $39.0057 \pm 3.5187$ & $55.2051 \pm 6.0904$ \\
Utilization $\uparrow$ & $56.93\% \pm 4.69\%$ & $50.69\% \pm 2.65\%$ & $45.20\% \pm 2.62\%$ & $41.93\% \pm 1.81\%$ & $40.20\% \pm 1.79\%$ \\
Resource Imbalance $\downarrow$ & $1.3891 \pm 0.3748$ & $1.6284 \pm 0.2536$ & $2.0353 \pm 0.2155$ & $2.0404 \pm 0.1813$ & $2.1362 \pm 0.2180$ \\
Distinct Resources & $13.39 \pm 0.93$ & $27.00 \pm 1.29$ & $34.86 \pm 1.27$ & $40.27 \pm 1.35$ & $44.01 \pm 1.94$ \\
Non-dominated Solutions $\uparrow$ & $32.50 \pm 3.64$ & $25.20 \pm 5.00$ & $22.10 \pm 4.16$ & $23.80 \pm 2.48$ & $21.20 \pm 4.38$ \\
Spacing (Schott's) $\downarrow$ & $0.0843 \pm 0.0255$ & $0.0998 \pm 0.0391$ & $0.0905 \pm 0.0241$ & $0.1029 \pm 0.0302$ & $0.1093 \pm 0.0258$ \\
Hypervolume $\uparrow$ & $0.7466 \pm 0.0531$ & $0.8634 \pm 0.1177$ & $0.9648 \pm 0.0769$ & $0.9979 \pm 0.0713$ & $0.9978 \pm 0.0826$ \\
\bottomrule
\end{tabular}
\end{table*}

\begin{table*}[htbp]
\centering
\caption{Impact of workflow depth $d(G)$ on NSGA-II scheduling quality (mean $\pm$ std over 20 independent workflows), $N=50$, $w(G)=8$ and all other parameters at their defaults.}
\label{table:workflow_depth}
\begin{tabular}{lrrrr}
\toprule
\textbf{Workflow depth $d(G)$} & \textbf{15} & \textbf{25} & \textbf{35} & \textbf{42} \\
\midrule
Makespan (s) $\downarrow$ & $2663.27 \pm 148.18$ & $4039.80 \pm 163.21$ & $5244.33 \pm 286.58$ & $6220.51 \pm 171.11$ \\
Cost (\$) $\downarrow$ & $6.1716 \pm 1.3416$ & $5.5210 \pm 1.0050$ & $4.8848 \pm 0.8934$ & $4.8246 \pm 0.5016$ \\
Utilization $\uparrow$ & $50.69\% \pm 2.65\%$ & $47.99\% \pm 2.07\%$ & $45.98\% \pm 3.59\%$ & $45.94\% \pm 1.85\%$ \\
Resource Imbalance $\downarrow$ & $1.6284 \pm 0.2536$ & $1.7264 \pm 0.2538$ & $1.9855 \pm 0.2214$ & $1.9307 \pm 0.2897$ \\
Distinct Resources & $27.00 \pm 1.29$ & $26.43 \pm 1.16$ & $26.36 \pm 1.03$ & $25.77 \pm 1.50$ \\
Non-dominated Solutions $\uparrow$ & $25.20 \pm 5.00$ & $28.00 \pm 5.08$ & $32.30 \pm 4.15$ & $33.50 \pm 3.23$ \\
Spacing (Schott's) $\downarrow$ & $0.0998 \pm 0.0391$ & $0.0781 \pm 0.0301$ & $0.0724 \pm 0.0220$ & $0.0800 \pm 0.0276$ \\
Hypervolume $\uparrow$ & $0.8634 \pm 0.1177$ & $0.8083 \pm 0.0588$ & $0.7482 \pm 0.0612$ & $0.7902 \pm 0.0967$ \\
\bottomrule
\end{tabular}
\end{table*}

\begin{table*}[htbp]
\centering
\caption{Impact of workflow width $w(G)$ on NSGA-II scheduling quality (mean $\pm$ std over 20 independent workflows), $N=50$, $d(G)=15$ and all other parameters at their defaults.}
\label{table:workflow_width}
\begin{tabular}{lrrrrr}
\toprule
\textbf{Workflow width $w(G)$} & \textbf{5} & \textbf{8} & \textbf{15} & \textbf{25} & \textbf{35} \\
\midrule
Makespan (s) $\downarrow$ & $2639.93 \pm 199.52$ & $2663.27 \pm 148.18$ & $2640.81 \pm 208.65$ & $2491.95 \pm 164.24$ & $2347.12 \pm 185.45$ \\
Cost (\$) $\downarrow$ & $6.2573 \pm 1.3518$ & $6.1716 \pm 1.3416$ & $4.7648 \pm 0.8270$ & $5.8199 \pm 2.4837$ & $6.5181 \pm 1.4381$ \\
Utilization $\uparrow$ & $48.15\% \pm 4.27\%$ & $50.69\% \pm 2.65\%$ & $50.09\% \pm 2.48\%$ & $55.94\% \pm 2.22\%$ & $59.09\% \pm 2.90\%$ \\
Resource Imbalance $\downarrow$ & $1.7168 \pm 0.2508$ & $1.6284 \pm 0.2536$ & $1.6345 \pm 0.2434$ & $1.6301 \pm 0.2838$ & $1.5288 \pm 0.2007$ \\
Distinct Resources & $26.73 \pm 1.87$ & $27.00 \pm 1.29$ & $25.24 \pm 1.78$ & $25.06 \pm 1.38$ & $24.19 \pm 2.27$ \\
Non-dominated Solutions $\uparrow$ & $24.80 \pm 5.00$ & $25.20 \pm 5.00$ & $21.80 \pm 3.06$ & $27.60 \pm 3.64$ & $28.10 \pm 4.53$ \\
Spacing (Schott's) $\downarrow$ & $0.0871 \pm 0.0266$ & $0.0998 \pm 0.0391$ & $0.1030 \pm 0.0334$ & $0.0927 \pm 0.0452$ & $0.0998 \pm 0.0379$ \\
Hypervolume $\uparrow$ & $0.9025 \pm 0.0924$ & $0.8634 \pm 0.1177$ & $0.8864 \pm 0.0909$ & $0.7647 \pm 0.1183$ & $0.7734 \pm 0.1053$ \\
\bottomrule
\end{tabular}
\end{table*}

\clearpage

\section{Cloud Resource Pool}
\label{sec:appendix_cloud_ablations}

This appendix provides tables for the impacts of cloud size and composition on the NSGA-II performance. 

\begin{table*}[htbp]
\centering
\small
\caption{Impact of cloud resource pool size on NSGA-II scheduling quality (mean $\pm$ std over 20 independent workflows).}
\label{table:cloud_size}
\begin{tabular}{lrrr}
\toprule
\textbf{Cloud size $|\mathcal{R}|$} & \textbf{60} & \textbf{180} & \textbf{300} \\
\midrule
Makespan (s) $\downarrow$ & $2663.27 \pm 148.18$ & $2676.70 \pm 65.61$ & $2591.85 \pm 142.67$ \\
Cost (\$) $\downarrow$ & $6.1716 \pm 1.3416$ & $17.8053 \pm 4.1071$ & $25.7325 \pm 3.6668$ \\
Utilization $\uparrow$ & $50.69\% \pm 2.65\%$ & $47.87\% \pm 1.95\%$ & $45.67\% \pm 2.15\%$ \\
Resource Imbalance $\downarrow$ & $1.6284 \pm 0.2536$ & $1.7417 \pm 0.1790$ & $1.8114 \pm 0.2025$ \\
Distinct Resources & $27.00 \pm 1.29$ & $36.39 \pm 1.40$ & $39.88 \pm 0.77$ \\
Non-dominated Solutions $\uparrow$ & $25.20 \pm 5.00$ & $27.00 \pm 4.60$ & $28.50 \pm 3.07$ \\
Spacing (Schott's) $\downarrow$ & $0.0998 \pm 0.0391$ & $0.0900 \pm 0.0292$ & $0.0839 \pm 0.0265$ \\
Hypervolume $\uparrow$ & $0.8634 \pm 0.1177$ & $0.8727 \pm 0.1149$ & $0.8284 \pm 0.1009$ \\
\bottomrule
\end{tabular}
\end{table*}

\begin{table*}[htbp]
\centering
\small
\caption{Impact of cloud resource pool composition on NSGA-II scheduling quality (mean $\pm$ std over 20 independent workflows).}
\label{table:resource_pool}
\begin{tabular}{lrrr}
\toprule
\textbf{Resource pool} & \textbf{CPU-only} & \textbf{GPU-only} & \textbf{CPU+GPU} \\
\midrule
Makespan (s) $\downarrow$ & $2774.33 \pm 157.20$ & $2620.16 \pm 124.50$ & $2663.27 \pm 148.18$ \\
Cost (\$) $\downarrow$ & $6.7696 \pm 1.4037$ & $64.7780 \pm 7.7081$ & $6.1716 \pm 1.3416$ \\
Utilization $\uparrow$ & $53.86\% \pm 3.33\%$ & $46.84\% \pm 3.13\%$ & $50.69\% \pm 2.65\%$ \\
Resource Imbalance $\downarrow$ & $1.4666 \pm 0.1971$ & $1.0554 \pm 0.1541$ & $1.6284 \pm 0.2536$ \\
Distinct Resources & $27.57 \pm 1.43$ & $22.25 \pm 1.37$ & $27.00 \pm 1.29$ \\
Non-dominated Solutions $\uparrow$ & $23.00 \pm 4.80$ & $15.60 \pm 2.76$ & $25.20 \pm 5.00$ \\
Spacing (Schott's) $\downarrow$ & $0.1003 \pm 0.0369$ & $0.1254 \pm 0.0423$ & $0.0998 \pm 0.0391$ \\
Hypervolume $\uparrow$ & $0.8166 \pm 0.1199$ & $1.0291 \pm 0.0481$ & $0.8634 \pm 0.1177$ \\
\bottomrule
\end{tabular}
\end{table*}

\clearpage

\bibliographystyle{unsrt}
\bibliography{reference}

\end{document}